\documentclass[
 reprint,
 amsmath,amssymb,
 aps,
 prd,
 preprintnumbers,
 superscriptaddress
]{revtex4-1}

\usepackage[
    colorlinks=true,
    linkcolor={red!50!black},
    urlcolor={green!50!black},
    citecolor={blue!80!black}
]{hyperref}

\usepackage{lmodern}
\usepackage{comment}

\usepackage[utf8]{inputenc}

\usepackage{amsmath,amsfonts,amssymb,amsthm}
\usepackage{mathtools}
\usepackage{mathrsfs}
\usepackage{bm}
\usepackage{tensor}

\newcommand{\mb}{\ensuremath{\mathbb}}
\newcommand{\mc}{\ensuremath{\mathcal}}
\newcommand{\mf}{\ensuremath{\mathfrak}}
\newcommand{\mh}{\ensuremath{\mathscr}}
\newcommand{\mr}{\ensuremath{\mathrm}}
\newcommand{\ms}{\ensuremath{\mathsf}}

\renewcommand{\v}[1]{\ensuremath{\bm{\mathbf{#1}}}}
\renewcommand{\d}{\mr{d}}

\newcommand{\C}{\mb{C}}
\renewcommand{\l}{\left}
\renewcommand{\r}{\right}
\newcommand{\f}{\frac}
\newcommand{\tf}{\tfrac}

\renewcommand{\t}{\text}

\newcommand{\I}{\indices}
\newcommand{\pd}[2]{\f{\partial{#1}}{\partial{#2}}}
\newcommand{\td}[2]{\f{\d{#1}}{\d{#2}}}

\DeclareMathOperator{\tr}{tr}
\DeclareMathOperator{\Pexp}{Pexp}
\DeclareMathOperator{\rk}{rk}

\newcommand{\hG}[5]{%
\tensor[_{#1}]{F}{_{#2}}%
\left(\genfrac..{0pt}{}{#3}{#4}\middle|#5%
\right)%
}

\makeatletter
\newcommand{\subalign}[1]{%
  \vbox{%
    \Let@ \restore@math@cr \default@tag
    \baselineskip\fontdimen10 \scriptfont\tw@
    \advance\baselineskip\fontdimen12 \scriptfont\tw@
    \lineskip\thr@@\fontdimen8 \scriptfont\thr@@
    \lineskiplimit\lineskip
    \ialign{\hfil$\m@th\scriptstyle##$&$\m@th\scriptstyle{}##$\crcr
      #1\crcr
    }%
  }
}
\makeatother

\usepackage[caption=false]{subfig}
\usepackage{tikz}

\begin{document}

\preprint{QMUL-PH-18-29}
\preprint{KIAS-P18108}

\title{Supersymmetric Wilson loops in two dimensions and duality}

\author{Rodolfo Panerai}
\email{r.panerai@qmul.ac.uk}
\affiliation{Centre for Research in String Theory, School of Physics and Astronomy,\\
Queen Mary University of London,\\
Mile End Road, E1 4NS London, United Kingdom}

\author{Matteo Poggi}
\email{mpoggi@kias.re.kr}
\affiliation{International School of Advanced Studies (SISSA) and INFN Sezione di Trieste,\\
via Bonomea 265, 34136 Trieste, Italy}
\affiliation{School of Physics, Korea Institute for Advanced Study (KIAS),\\
85 Hoegi-ro, Dongdaemun-gu, Seoul 02455, Republic of Korea
}

\author{Domenico Seminara}
\email{seminara@fi.infn.it}
\affiliation{Dipartimento di Fisica, Universit\`a di Firenze and INFN Sezione di Firenze,\\
via G. Sansone 1, 50019 Sesto Fiorentino, Italy}

\begin{abstract}
We classify bosonic $\mc{N}=(2,2)$ supersymmetric Wilson loops on arbitrary backgrounds with vector-like R-symmetry. These can be defined on any smooth contour and come in two forms which are universal across all backgrounds. We show that these Wilson loops, due to their cohomological properties, are all invariant under smooth deformations of their contour. At genus zero they can always be mapped to local operators and computed exactly with supersymmetric localisation. Finally, we find the precise map, under two-dimensional Seiberg-like dualities, of correlators of supersymmetric Wilson loops.
\end{abstract}

\maketitle

\section{Introduction}
Wilson loops are an important tool to understand the dynamics of gauge theories: they encode how an (infinitely massive) elementary excitation localised along the loop responds to the presence of a dynamical gauge field. They provide an efficient and operative way to characterise the confinement or deconfinement phases of the theory according to whether their expectation value increases with the area (\emph{area law}), or, respectively, with the perimeter of the loop (\emph{perimeter law}) \cite{Wilson:1974sk}.

In supersymmetric theories, a special role is played by the so-called supersymmetric or BPS Wilson loops. These observables are annihilated by a subsector of the supercharges of theory and they are often amenable to exact evaluation through localisation techniques.
The first and probably most famous example is the Maldacena-Wilson loop along a circular path in $\mathcal{N}=4$ super Yang--Mills (sYM) in four dimensions. It preserves half of the original supercharges and its expectation value is given by a Gaussian matrix model whose form was  originally conjectured  in \cite{Erickson:2000af} and then rigorously derived by Pestun in \cite{Pestun:2007rz}. According to the AdS/CFT correspondence, the very same quantity can be  computed at strong coupling by a semiclassical string computation, the actual matching at leading \cite{Erickson:2000af} and subleading \cite{Kruczenski:2008zk,Forini:2015bgo,Medina-Rincon:2018wjs} order being in striking support for the gauge/gravity paradigm.

This initial success has prompted an intensive search for other examples both in $\mathcal{N}=4$ sYM in four dimensions and in other theories with a different number of supersymmetries or dimensions. In \cite{Zarembo:2002an}, Zarembo constructed entire families of BPS Wilson loops in $\mc{N}=4$ sYM that preserve $\f{1}{16}$, $\f{1}{8}$ or $\f{1}{4}$ of the supersymmetry depending on the subspace spanned by the contour supporting the loop. The shape of the contour is not relevant for the existence of unbroken supercharges. Unfortunately, the expectation value of these observable does not receive quantum correction and it is one to all order in the coupling constant. Nonetheless, this triviality carries significant information about the dynamics of the theory especially in the context of the AdS/CFT correspondence: the associated family of minimal surfaces in $\mr{AdS}_5\times S^5$ must have zero regularized area independently of the shape of the boundary \cite{Dymarsky:2006ve}.

On the other hand, generic contours on an $S^3$ embedded in $\mathbb{R}^4$ and preserving at least two supercharges were constructed in \cite{Drukker:2007qr}. Restricting the loop on $S^2$, the authors conjectured that this can be computed in terms of the zero-instanton sector of similar observables in purely bosonic Yang--Mills in two dimensions on the sphere. This conjecture was first verified perturbatively up to two loops in \cite{Bassetto:2008yf,Young:2008ed} and later an (almost) complete proof using localisation techniques was given in \cite{Pestun:2009nn} (see also \cite{Bonini:2015fng}). The investigation of this family of Wilson loops has led to exact results for correlators of Wilson loops \cite{Bassetto:2009rt,Bassetto:2009ms,Giombi:2009ds,Giombi:2012ep}, correlators of Wilson loops and primary operators and theories living on one-dimensional defects \cite{Giombi:2018qox,Liendo:2018ukf,Giombi:2018hsx}. The defect approach was also crucial in relating circular Wilson loops to the energy emitted by an accelerating particle, captured by the so-called \emph{Bremsstrahlung function} \cite{Correa:2012at}.

In the $\mc{N}=2$ case, the original analysis by Pestun carries over and the $\f{1}{2}$-BPS circular Wilson loop has been studied: the relevant matrix model is much more complicated and the eigenvalue measure depends also on instanton contributions. The large-$N$ limit, where instantons decouple, has been thoroughly studied and successfully compared with AdS/CFT predictions \cite{Passerini:2011fe,Buchel:2013id,Russo:2013qaa}. Furthermore, also in this case an exact prediction for the Bremsstrahlung function has been put forward \cite{Fiol:2015spa, Bianchi:2018zpb}.

In three dimensions the situation is more intricate especially when considering superconformal theories. One can construct bosonic Wilson loops, whose structure mimics the behaviour in four dimensions: the connection appearing in the holonomy is built by adding to the gauge field a suitable bilinear in the scalars of the theory. These loops are supersymmetric only for a suitable choice of the contour (for instance circles and straight-lines) \cite{Drukker:2008zx,Chen:2008bp}. The expectation value of the circular bosonic Wilson loop can be again evaluated with a localisation procedure, developed by Kapustin, Willett and Yaakov \cite{Kapustin:2009kz}, and can be in principle computed for a very general class of $\mathcal{N}=2$ theories on $S^3$ (see \cite{Farquet:2014bda} for extensions on more general three-dimensional manifolds) through complicated matrix models.

In the superconformal case of ABJ(M) theories, the above Wilson loop has been explicitly evaluated by a tractable matrix model, which is closely related to that describing topological Chern--Simons theory on lens space \cite{Marino:2009jd,Drukker:2010nc}. However, in the context of the AdS/CFT correspondence, these bosonic loops are not dual to the fundamental strings since they in general possess the wrong residual symmetries. The holographic dual for the case of the straight line and the circle in ABJ(M) theories was constructed by Drukker and Trancanelli in \cite{Drukker:2009hy} and quite surprisingly couples, in addition to the gauge and scalar fields of the theory, also to the fermions in the bi-fundamental representation of the $\mr{U}(N) \times \mr{U}(M)$ gauge group. In other words, the usual connection is replaced by a superconnection built out of the fundamental fields and  living in the superalgebra $\mr{U}(N|M)$. The presence of fermionic couplings naturally allows for a larger number of preserved supersymmetries. These loops were extended to more general contours in \cite{Cardinali:2012ru} and to theories with less supersymmetries in \cite{Cooke:2015ila,Ouyang:2015iza,Ouyang:2015bmy,Lietti:2017gtc,Mauri:2018fsf}. They also allowed, as in four dimension, for the exact computation of the Bremsstrahlung function \cite{Bianchi:2014laa,Bianchi:2017ozk,Bianchi:2018scb,Bianchi:2018bke}

In two dimensions the situation is expected to be somewhat simpler, but it is still much less explored nonetheless. Interestingly, exact and very non trivial results can be obtained even for Wilson loops in non supersymmetric models, when the theory is defined on compact two-dimensional manifolds. For instance, in pure Yang--Mills a generic Wilson loop can be evaluated on any Riemann surface by exploiting either lattice  \cite{Rusakov:1990rs} or localisation \cite{Witten:1992xu} techniques. At large $N$ they exhibit stringy behavior \cite{Gross:1993hu,Gross:1993yt}, different scalings charting the phase structure \cite{Douglas:1993iia} and admit explicit solutions of the Migdal--Makeenko equations \cite{Kazakov:1980zi,Kazakov:1980zj}. 

It is quite natural therefore to wonder if a similar variety of phenomena is shared by the supersymmetric analogue of the ``canonical'' Wilson loops of two-dimensional Yang--Mills. More generally, one would like to classify and, hopefully, compute new gauge invariant observables in two-dimensional supersymmetric theories that could be useful in checking the AdS/CFT correspondence, testing non-perturbative dualities and constructing defect field theories. We focus in particular on $\mc{N}=(2,2)$ gauge theories with vector R-symmetry. Backgrounds for such theories on arbitrary Riemann surfaces were recently studied in \cite{Closset:2014pda,Bae:2015eoa}. More specifically, we would like to classify all bosonic Wilson loops built from the gauge supermultiplet which preserve some supersymmetries independently of the shape of the closed contour, realizing a sort of two-dimensional version of the Zarembo's charting in four-dimensions \cite{Zarembo:2002an}. We find two families of BPS observables 
\begin{align}\label{EQ:WL_intro}
W_{\epsilon} &= \tr \Pexp \int \mr{i}(A^{\ms{a}} + f_{\epsilon}^{\ms{a}}\sigma + \tilde{f}_{\epsilon}^{\ms{a}}\tilde{\sigma}) \, \dot{x}^{\ms{a}} \, \d t \;, \cr
W_{\tilde{\epsilon}} &= \tr \Pexp \int \mr{i}(A^{\ms{a}} + f_{\tilde{\epsilon}}^{\ms{a}}\sigma + \tilde{f}_{\tilde{\epsilon}}^{\ms{a}}\tilde{\sigma}) \, \dot{x}^{\ms{a}} \, \d t \;, 
\end{align}
where $\sigma$ and $\tilde{\sigma}$ are the scalar fields in the gauge supermultiplet and the couplings $f^{\ms{a}}$ and $\tilde{f}^{\ms{a}}$ are defined in \eqref{EQ:fL_definition} and \eqref{EQ:fR_definition}. These loops are $\f{1}{4}$-BPS and the analysis of the preserved supersymmetries extends to the case of a general Riemann surface.
For both classes of Wilson loops, we find that the associated field strength is $\v{Q}$-exact. This, in turn, implies that the quantum expectation value of a generic non-self-intersecting loop does not change under smooth deformations of its contour, thus signaling a topological character of the observable.

We then proceed to extract exact expectation values and correlators of these Wilson loops. For $\mc{N}=(2,2)$ gauge theories, exact results on various backgrounds have been computed in recent years using localisation. On the sphere \cite{Doroud:2012xw,Benini:2012ui}, these can be represented both as a sum over topological sectors of a matrix integral over the Cartan subalgebra of the gauge group (the so-called Coulomb-branch representation) and as the product of a vortex times an antivortex partition function, weighted by semiclassical factors and summed over isolated points on the Higgs branch. This dual representation is reminiscent of the non-supersymmetric case, where the exact partition function can be expressed both as a sum over the irreducible representations of the gauge group \cite{Rusakov:1990rs} and as a weighted sum over instanton solutions \cite{Witten:1992xu}. Fayet--Iliopoulos and theta terms for the abelian factors of the gauge group and twisted masses and background fluxes for the chiral multiplets can also be added, enriching the parametric dependence of the results. In this theory a one-parameter family of Wilson loops with contours lying on latitudes of the round two-sphere were already considered in \cite{Doroud:2012xw}: on the maximal circle the dimensionally reduced $\f{1}{6}$-BPS Wilson loop of three-dimensional $\mc{N}=2$ theories is recovered and the quantum expectation value of whole family is independent from the latitude angle.

These are recovered in our construction as the limit case in which the two Wilson loops in \eqref{EQ:WL_intro} coincide and the preserved supersymmetry is enhanced to $\f{1}{2}$-BPS. It is only in this case that the matrix model of \cite{Doroud:2012xw,Benini:2012ui} can be directly applied to perform exact computations. Our cohomological argument outlined earlier, however, ensures that the result can be extended to encompass Wilson loops on arbitrary contours. The final result can be expressed both in the Coulomb branch representation and in the Higgs branch representation: the non-triviality of the vacuum expectation value is ensured by the presence of twisted masses or background fluxes.

Two-dimensional gauge theories with $\mc{N}=(2,2)$ supersymmetry admit dual descriptions in the infrared regime, first described by Hori and Tong \cite{Hori:2006dk, Hori:2011pd}, analogously to the case of four-dimensional $\mc{N}=1$ Seiberg duality \cite{Seiberg:1994pq}. For unitary and special-unitary groups, the duality has been explicitly checked, at the level of the partition functions, using results from localisation \cite{Benini:2012ui,Doroud:2012xw}. 
More recently, the duality has been tested against correlators of Coulomb-branch operators for topologically-twisted theories \cite{Closset:2017vvl}.
Here, we show that the dictionary for unitary groups can be extended to include correlators of the Wilson loop operators in \eqref{EQ:WL_intro}, similarly to what has been done in \cite{Kapustin:2013hpk} for three-dimensional $\mc{N}=2$ theories. We find that a single operator insertion is mapped to a combination of Wilson loops in different representations.

The structure of the paper is the following: in Section~\ref{SEC:Wilson lines} we discuss the general construction of supersymmetric Wilson lines for two-dimensional $\mc{N}=(2,2)$ supersymmetric gauge theories with $\mr{U}(1)_{\mr{R}}$ vector R-symmetry on orientable Riemann surfaces. We derive the explicit form of the scalar couplings in term of the relevant Killing spinors and we are able to associate to a generic curve two $\f{1}{4}$-BPS Wilson lines. By further constraining the contours we observe a collapse of the two solutions into a single $\f{1}{2}$-BPS Wilson line. We then look at specific backgrounds and derive the explicit form of these $\f{1}{2}$-BPS lines.
Section~\ref{SEC:WL} concerns Wilson loops and the main result of the paper is presented: non-self-intersecting Wilson loops whose paths are homotopically equivalent are $\v{Q}$-cohomologous. In other words, the vacuum expectation value of our $\f{1}{4}$-BPS observables does not change under smooth deformations of their contour. We will prove this property by showing that the effect of an infinitesimal deformation results into a $\v{Q}$-exact quantity. In Section~\ref{SEC:localisation} we specialise to a squashed background on $S^2$ and find exact results for the value of any of these Wilson loops and their correlators, using supersymmetric localisation. We then discuss the dependence on geometrical data and external parameters. Finally, in Section~\ref{SEC:dualities} we determine how correlators of Wilson loops are mapped under Seiber-like dualities and provide explicit examples.

\section{Wilson lines}
\label{SEC:Wilson lines}
\subsection{\texorpdfstring{$\mc{N}=(2,2)$}{N=(2,2)} supersymmetry}\label{SEC:supersymmetry}
Let us consider two-dimensional $\mc{N}=(2,2)$ supersymmetric gauge theories with $\mr{U}(1)_{\mr{R}}$ vector R-symmetry on some orientable Riemann surface $\mh{M}$. Any supersymmetric background for such theories can be understood as a background off-shell supergravity multiplet as in the approach of \cite{Festuccia:2011ws}. In addition to the metric $g$, the multiplet contains, as bosonic degrees of freedom, a gauge field $B$, which couples to the R-symmetry current, and graviphotons $C$ and $\tilde{C}$, which couple to the central-charge currents. The Killing spinor equations for the generators of rigid supersymmetry are obtained by imposing the variation of the gravitini to be vanishing. This gives
\begin{align}\label{EQ:Killing_spinor_equations}
    (\nabla-\mr{i}B)\,\epsilon &= -\tf{1}{2}e^\ms{a}\gamma^\ms{a}(\tilde{H}\mr{P}_++H\mr{P}_-)\,\epsilon \;, \cr
    (\nabla+\mr{i}B)\,\tilde{\epsilon} &= -\tf{1}{2}e^\ms{a}\gamma^\ms{a}(H\mr{P}_++\tilde{H}\mr{P}_-)\,\tilde{\epsilon} \;,
\end{align}
where $H$ and $\tilde{H}$ are the dual field strengths of the graviphotons
\begin{align}
    H &= -\tf{\mr{i}}{2} \, {*}\d C \;,
    &
    \tilde{H} &= -\tf{\mr{i}}{2} \, {*}\d \tilde{C} \;.
\end{align}
The various backgrounds induced by the solutions of above equation on an arbitrary $\mh{M}$ have been studied an classified in \cite{Closset:2014pda,Bae:2015eoa}. The only backgrounds that apply to any genus $\v{g}$ are the topological A- and $\bar{\mr{A}}$-twist \cite{Witten:1988xj,Eguchi:1990vz}, which preserve up to two supercharges of opposite R-charge. All other possible backgrounds apply to the case of $\v{g}<2$. For $\v{g}=1$ one can adopt a flat background as on the complex plane $\C$. Finally, the case of $\v{g}=0$ admits various rigid supersymmetry realisations preserving a different number of supercharges. We refer the reader to \cite{Closset:2014pda,Bae:2015eoa} for more detail. In this paper, we will pay particular attention to the maximally-supersymmetric round sphere and its squashing.

While still keeping the supersymmetric background generic, let us look at $\mc{N}=(2,2)$ vector multiplets. Here we denote with $\v{G}$ the gauge group and with $\mf{g}$ its associated Lie algebra. In the Wess--Zumino gauge the multiplet consists of $\mf{g}$-valued fields $(A,\lambda,\tilde{\lambda},\sigma,\tilde{\sigma},D)$, where $A$ is the gauge vector field, $\lambda$ and $\tilde{\lambda}$ are gaugini, $\sigma$ and $\tilde{\sigma}$ are scalars and $D$ is an auxiliary scalar field. The $\mr{U}(1)_{\mr{R}}$-charges for the component fields are given by
\[
    \begin{tabular}{c @{\quad} c @{\quad} c @{\quad} c @{\quad} c @{\quad} c @{\quad} c @{\quad} c}
    \toprule
    $A$ & $\lambda^+$ & $\lambda^-$ & $\tilde{\lambda}^+$ & $\tilde{\lambda}^-$ & $\sigma$ & $\tilde{\sigma}$ & $D$ \\
    \colrule
    $0$ & $+1$ & $+1$ & $-1$ & $-1$ & $0$ & $0$ & $0$ \\
    \botrule
    \end{tabular}
\]
In general the axial R-symmetry is explicitly broken. The supersymmetric variations for the component fields are given in Appendix \ref{APP:supersymmetry}.

\subsection{Supersymmetric connections}
We are now ready to introduce a generic Wilson line of the form
\begin{align}
W = \Pexp \int_{\Gamma} \mr{i}\mc{A} \;,
\end{align}
where $\Gamma\colon(0,1)\to\mh{M}$ is some smooth path and $\mc{A}$ is a $\v{G}$-connection defined as some combination of fields from the vector multiplet. In coordinates, we denote $\Gamma$ as the embedding $x(t)$.

Since we are interested in bosonic deformations of the non-supersymmetric Wilson line defined with $\mc{A}=A$, we write
\begin{align}\label{EQ:Wcal}
\mc{A} = A + f\sigma + \tilde{f}\tilde{\sigma} \;,
\end{align}
for some one-forms $f$ and $\tilde{f}$.

We want to find the choices of $f$ and $\tilde{f}$ for which $W$ is annihilated by some supercharge. For this, we need the explicit expression of the supersymmetric variation of the fields $A$, $\sigma$ and $\tilde{\sigma}$. Interestingly, these fields are precisely those whose variations are insensitive to the particular realisation of supersymmetry considered, as can be seen from \eqref{EQ:vector_multiplet_variations}.

When acting with a supersymmetry variation on $\mc{A} = \mc{A}^{\ms{a}} e^{\ms{a}}$, one finds
\begin{align}
\delta\mc{A}^{\ms{a}} = \epsilon M^{\ms{a}} \tilde{\lambda} + \tilde{\epsilon} \tilde{M}^{\ms{a}} \lambda \;,
\end{align}
where
\begin{align}
M^{\ms{a}} &= + \tf{\mr{i}}{2}\gamma^{\ms{a}} + f^{\ms{a}}\mr{P}_+ + \tilde{f}^{\ms{a}}\mr{P}_- \;, \cr
\tilde{M}^{\ms{a}} &= - \tf{\mr{i}}{2}\gamma^{\ms{a}} + f^{\ms{a}}\mr{P}_- + \tilde{f}^{\ms{a}}\mr{P}_+ \;.
\end{align}

From this, one can immediately obtain the variation of the integrand $\mc{A}^{\ms{a}}\dot{x}^{\ms{a}}$, i.e.\ of the pullback of $\mc{A}$ on the path $\Gamma$.
The kernels of both $M^\ms{a}\dot{x}^{\ms{a}}$ and $\tilde{M}^{\ms{a}}\dot{x}^{\ms{a}}$ are nontrivial for any choice of $x(t)$ if
\begin{align}
f^{\ms{1}} \tilde{f}^{\ms{1}} = f^{\ms{2}} \tilde{f}^{\ms{2}} = -\tf{1}{4} \;, \qquad f^{\ms{1}} \tilde{f}^{\ms{2}} + f^{\ms{2}} \tilde{f}^{\ms{1}} = 0 \;.
\end{align}
When these conditions are satisfied, one can find solutions for $f$ and $\tilde{f}$ such that the Wilson line is annihilated by either $\v{Q} = \epsilon Q$ for
\begin{align}\label{EQ:fL_definition}
f_{\epsilon} &= +\f{\mr{i}}{2} \f{\epsilon^+}{\epsilon^-} (e^{\ms{1}} + \mr{i}e^{\ms{2}}) \;, & \tilde{f}_{\epsilon} &= +\f{\mr{i}}{2} \f{\epsilon^-}{\epsilon^+} (e^{\ms{1}} - \mr{i}e^{\ms{2}}) \;,
\end{align}
or by $\tilde{\v{Q}} = \tilde{\epsilon} \tilde{Q}$ for
\begin{align}\label{EQ:fR_definition}
f_{\tilde{\epsilon}} &= - \f{\mr{i}}{2} \f{\tilde{\epsilon}^-}{\tilde{\epsilon}^+} (e^{\ms{1}} - \mr{i}e^{\ms{2}}) \;, & \tilde{f}_{\tilde{\epsilon}} &= -\f{\mr{i}}{2} \f{\tilde{\epsilon}^+}{\tilde{\epsilon}^-} (e^{\ms{1}} + \mr{i}e^{\ms{2}}) \;,
\end{align}
for any choice of the path $\Gamma$.

Notice how the above are well defined: the ratios of spin components are single-valued on $\mh{M}$ and, under a frame change, transform with a phase that cancels the opposite phase coming from the holomorphic/anti-holomorphic combinations of zweibein elements.

To summarise our discussion so far, given a generic path $x(t)$ we have defined two $\f{1}{4}$-BPS Wilson lines, namely
\begin{align}
W_{\epsilon} &= \Pexp \int \mr{i}\mc{A}_{\epsilon}^{\ms{a}} \dot{x}^{\ms{a}} \, \d t \;, & \mc{A}_{\epsilon}^{\ms{a}} &= A^{\ms{a}} + f_{\epsilon}^{\ms{a}}\sigma + \tilde{f}_{\epsilon}^{\ms{a}}\tilde{\sigma} \;, \label{EQ:W_L} \\
W_{\tilde{\epsilon}} &= \Pexp \int \mr{i}\mc{A}_{\tilde{\epsilon}}^{\ms{a}} \dot{x}^{\ms{a}} \, \d t \;, & \mc{A}_{\tilde{\epsilon}}^{\ms{a}} &= A^{\ms{a}} + f_{\tilde{\epsilon}}^{\ms{a}}\sigma + \tilde{f}_{\tilde{\epsilon}}^{\ms{a}}\tilde{\sigma} \;. \label{EQ:W_R}
\end{align}
annihilated, respectively, by $\v{Q}$ and $\tilde{\v{Q}}$.
We want to stress that the above construction is fully general and, following from our previous discussion, holds for any supersymmetric background, provided that this preserves the selected supercharge $\v{Q}$ or $\tilde{\v{Q}}$.

For certain choices of $\Gamma$, the two Wilson lines may coincide. This happens when $x(t)$ obeys the differential equation
\begin{align}\label{EQ:1/2_BPS_equation}
\f{\epsilon^-}{\epsilon^+} \f{\tilde{\epsilon}^-}{\tilde{\epsilon}^+} = -\f{\dot{x}^{\ms{1}}+\mr{i}\dot{x}^{\ms{2}}}{\dot{x}^{\ms{1}}-\mr{i}\dot{x}^{\ms{2}}} \;.
\end{align}
On such paths, we obtain a unique Wilson line which is $\f{1}{2}$-BPS, as it is annihilated both by $\v{Q}$ and $\tilde{\v{Q}}$. From \eqref{EQ:1/2_BPS_equation}, a necessary condition on the supercharges is imposed, namely
\begin{align}
\bigg|\f{\epsilon^-}{\epsilon^+} \f{\tilde{\epsilon}^-}{\tilde{\epsilon}^+}\bigg|^2 = 1 \;.
\end{align}

The Killing spinor equations fix the action of the exterior derivative of $f$ and $\tilde{f}$. From \eqref{EQ:Killing_spinor_equations}, in fact, follow
\begin{align}
\d\epsilon^+ &= -\tf{1}{2}H\l(e^{\ms{1}}-\mr{i}e^{\ms{2}}\r) \epsilon^- - \l(\tf{1}{2}\omega^{\ms{12}} + \mr{i}B\r) \epsilon^+ \;, \cr
\d\epsilon^- &= -\tf{1}{2}\tilde{H}\l(e^{\ms{1}}+\mr{i}e^{\ms{2}}\r) \epsilon^+ + \l(\tf{1}{2}\omega^{\ms{12}} - \mr{i}B\r) \epsilon^- \;, \cr
\d\tilde{\epsilon}^+ &= -\tf{1}{2}\tilde{H}\l(e^{\ms{1}}-\mr{i}e^{\ms{2}}\r) \tilde{\epsilon}^- - \l(\tf{1}{2}\omega^{\ms{12}} - \mr{i}B\r) \tilde{\epsilon}^+ \;, \cr
\d\tilde{\epsilon}^- &= -\tf{1}{2}H\l(e^{\ms{1}}+\mr{i}e^{\ms{2}}\r) \tilde{\epsilon}^+ + \l(\tf{1}{2}\omega^{\ms{12}} + \mr{i}B\r) \tilde{\epsilon}^- \;.
\end{align}
When using the above with \eqref{EQ:fL_definition} and \eqref{EQ:fR_definition}, the dependence on $\omega$ and $B$ drops and one is left with
\begin{align}
    \d f_{\epsilon} &= \d f_{\tilde{\epsilon}} = +\tf{1}{2}H \, e^{\ms{1}} \wedge e^{\ms{2}} \;, \cr
    \d \tilde{f}_{\epsilon} &= \d \tilde{f}_{\tilde{\epsilon}} = -\tf{1}{2}\tilde{H} \, e^{\ms{1}} \wedge e^{\ms{2}} \;.
\end{align}

Starting from the supersymmetric connections $\mc{A}_{\epsilon}$ and $\mc{A}_{\tilde{\epsilon}}$, we can introduce the associated field strengths
\begin{align}
\mc{F} &= \d\mc{A} - \mr{i}\mc{A}\wedge\mc{A} \;.
\end{align}
The dual forms read
\begin{align}\label{EQ:Qexactness}
*\mc{F} = \;&{*}F - \mr{i}\varepsilon^{\ms{ab}}f^{\ms{a}}\tilde{f}^{\ms{b}}[\sigma,\tilde{\sigma}] + \varepsilon^{\ms{ab}}(f^{\ms{b}}\mr{D}^{\ms{a}}\sigma + \tilde{f}^{\ms{b}}\mr{D}^{\ms{a}}\tilde{\sigma}) \cr
&+ *\d f \, \sigma + *\d\tilde{f} \, \tilde{\sigma} \;.
\end{align}
Here we have used that fact $A$, $\sigma$ and $\tilde\sigma$ are uncharged with respect to R-symmetry.

One crucial property of $\mc{F}_{\epsilon}$ and $\mc{F}_{\tilde{\epsilon}}$ is that they are respectively $\v{Q}$- and $\tilde{\v{Q}}$-exact. In particular,
\begin{align}\label{EQ:F_exact}
*\mc{F}_{\epsilon} &= -\f{\v{Q}(\epsilon\lambda)}{2\epsilon^+\epsilon^-} \;, &
*\mc{F}_{\tilde{\epsilon}} &= +\f{\tilde{\v{Q}}(\tilde{\epsilon}\tilde{\lambda})}{2\tilde{\epsilon}^+\tilde{\epsilon}^-} \;.
\end{align}
The above can be obtained by using the identities
\begin{align}
 -\mr{i}\f{\epsilon\mr{P}_-\gamma^{\ms{a}}\epsilon}{2\epsilon^+\epsilon^-} &= \varepsilon^{\ms{ab}}f_{\epsilon}^{\ms{b}} \;, & 
 -\mr{i}\f{\epsilon\mr{P}_+\gamma^{\ms{a}}\epsilon}{2\epsilon^+\epsilon^-} &= \varepsilon^{\ms{ab}}\tilde{f}_{\epsilon}^{\ms{b}} \;,
 \cr
 \mr{i}\f{\tilde{\epsilon}\mr{P}_+\gamma^{\ms{a}}\tilde{\epsilon}}{2\tilde{\epsilon}^+\tilde{\epsilon}^-} &= \varepsilon^{\ms{ab}}f_{\tilde{\epsilon}}^{\ms{b}} \;, &
 \mr{i}\f{\tilde{\epsilon}\mr{P}_-\gamma^{\ms{a}}\tilde{\epsilon}}{2\tilde{\epsilon}^+\tilde{\epsilon}^-} &= \varepsilon^{\ms{ab}}\tilde{f}_{\tilde{\epsilon}}^{\ms{b}} \;,
\end{align}
and
\begin{align}
 -\mr{i}\f{\epsilon\mr{P}_-\gamma^{\ms{a}}\mr{D}^{\ms{a}}\epsilon}{2\epsilon^+\epsilon^-} &= \mr{i}\f{\tilde{\epsilon}\mr{P}_+\gamma^{\ms{a}}\mr{D}^{\ms{a}}\tilde{\epsilon}}{2\tilde{\epsilon}^+\tilde{\epsilon}^-} = *\d f_{\epsilon} = *\d f_{\tilde{\epsilon}} \;, \cr
 -\mr{i}\f{\epsilon\mr{P}_+\gamma^{\ms{a}}\mr{D}^{\ms{a}}\epsilon}{2\epsilon^+\epsilon^-} &= \mr{i}\f{\tilde{\epsilon}\mr{P}_-\gamma^{\ms{a}}\mr{D}^{\ms{a}}\tilde{\epsilon}}{\tilde{\epsilon}^+\tilde{\epsilon}^-} = *\d \tilde{f}_{\epsilon} = *\d \tilde{f}_{\tilde{\epsilon}} \;.
\end{align}
This fact will play a prominent role in Section~\ref{SEC:WL}.

In the remainder of this section we will look at some specific backgrounds.

\subsection{Flat backgrounds}
We begin by studying Wilson lines on manifolds that admit a flat background, i.e.\ the plane, the cylinder and the torus. On a flat background, Killing spinors are constant. This, in turn, implies that $f^{\ms{a}}$ and $\tilde{f}^{\ms{a}}$ are constant for both choices of Wilson lines $W_{\epsilon}$ and $W_{\tilde{\epsilon}}$. The paths associated with $\f{1}{2}$-BPS Wilson lines are straight lines, as it can be seen by solving \eqref{EQ:1/2_BPS_equation} for constant $\epsilon$ and $\tilde{\epsilon}$.

\subsection{The round sphere}\label{SEC:sphere}
Let us now consider the case of $\mh{M} \simeq S^2$ equipped with the round metric. We denote the radius of the sphere with $r$ and use spherical coordinates $\theta\in[0,\pi]$ and $\varphi\in[0,2\pi)$ in terms of which the zweibein reads
\begin{align}
e^{\ms{1}} &= r \, \d\theta\;, \cr
e^{\ms{2}} &= r\sin\theta \, \d\varphi \;.
\end{align}
We follow \cite{Benini:2012ui} and consider the supersymmetry algebra generated by Killing spinors obeying \eqref{EQ:Killing_spinor_equations} with
\begin{align}
    B&=0 \;, & H=\tilde{H}=-\mr{i}/r \;.
\end{align}
A generic solution is of the form
\begin{align}
\epsilon &= e^{\mr{i}\f{\theta}{2}\gamma^1} \, e^{\mr{i}\f{\varphi}{2}\gamma^3} \epsilon_0 \;, \cr
\tilde{\epsilon} &= e^{\mr{i}\f{\theta}{2}\gamma^1} \, e^{\mr{i}\f{\varphi}{2}\gamma^3} \tilde{\epsilon}_0 \;,
\end{align}
for some constant spinors $\epsilon_0$ and $\tilde{\epsilon}_0$.

It will prove to be useful to also work in terms of the complex stereographic projection
\begin{align}
z &= 2r \tan\tf{\theta}{2} \, e^{i\varphi} \;, \cr
\bar{z} &= 2r \tan\tf{\theta}{2} \, e^{-i\varphi} \;.
\end{align}
With this choice of variables, we find
\begin{align}
f_{\epsilon} &= +\f{\mr{i}}{2} \l(1+\f{z\bar{z}}{4r^2}\r)^{-1} \f{2r\epsilon_0^{+} + \mr{i}\epsilon_0^{-}\bar{z}}{2r\epsilon_0^{-} + \mr{i}\epsilon_0^{+}z} \; \d z \;, \cr
\tilde{f}_{\epsilon} &= +\f{\mr{i}}{2} \l(1+\f{z\bar{z}}{4r^2}\r)^{-1} \f{2r\epsilon_0^{-} + \mr{i}\epsilon_0^{+}z}{2r\epsilon_0^{+} + \mr{i}\epsilon_0^{-}\bar{z}} \; \d\bar{z} \;, \cr
f_{\tilde{\epsilon}} &= -\f{\mr{i}}{2} \l(1+\f{z\bar{z}}{4r^2}\r)^{-1} \f{2r\tilde{\epsilon}_0^{-} + \mr{i}\tilde{\epsilon}_0^{+}z}{2r\tilde{\epsilon}_0^{+} + \mr{i}\tilde{\epsilon}_0^{-}\bar{z}} \; \d \bar{z} \cr
\tilde{f}_{\tilde{\epsilon}} &= -\f{\mr{i}}{2} \l(1+\f{z\bar{z}}{4r^2}\r)^{-1} \f{2r\tilde{\epsilon}_0^{+} + \mr{i}\tilde{\epsilon}_0^{-}\bar{z}}{2r\tilde{\epsilon}_0^{-} + \mr{i}\tilde{\epsilon}_0^{+}z} \; \d z \;.
\end{align}

We see that $f_{\epsilon}$ has a unique singular point at $z_{\epsilon,\mr{1}} = 2\mr{i}r \epsilon_0^-/\epsilon_0^+$ and a unique zero at \mbox{$z_{\epsilon,\mr{2}} = -2\mr{i}r \bar{\epsilon}_0^+/\bar{\epsilon}_0^-$}. The same is true for $\tilde{f}_{\epsilon}$, but the locations of the singularity and the zero are swapped. Moreover, the fact that $z_{\epsilon,\mr{1}}\bar{z}_{\epsilon,\mr{2}} = -4r^2$ implies that the two points on $S^2$ are antipodal.
Notice how here the components of $\epsilon_0$ play the role of homogeneous coordinates.

The case of $f_{\tilde{\epsilon}}$ and $\tilde{f}_{\tilde{\epsilon}}$ is completely analogous: $f_{\tilde{\epsilon}}$ has a singularity at $z_{\tilde{\epsilon},\mr{1}} = -2\mr{i}r \bar{\tilde{\epsilon}}_0^+/\bar{\tilde{\epsilon}}_0^-$ and a zero at the antipodal point $z_{\tilde{\epsilon},\mr{2}} = 2\mr{i}r \tilde{\epsilon}_0^-/\tilde{\epsilon}_0^+$, whether the converse is true for $\tilde{f}_{\tilde{\epsilon}}$.

The paths $z(t)$ over which the Wilson line becomes $\f{1}{2}$-BPS satisfy, according to \eqref{EQ:1/2_BPS_equation},
\begin{align}\label{EQ:1/2_BPS_equation_on_S2}
0 = \;&\f{\dot{z}}{(2r\epsilon_0^{-} + \mr{i}\epsilon_0^{+}z)(2r\tilde{\epsilon}_0^{-} + \mr{i}\tilde{\epsilon}_0^{+}z)} \cr
&+ \f{\dot{\bar{z}}}{(2r\epsilon_0^{+} + \mr{i}\epsilon_0^{-}\bar{z})(2r\tilde{\epsilon}_0^{+} + \mr{i}\tilde{\epsilon}_0^{-}\bar{z})} \;.
\end{align}
Suppose that $\det(\epsilon_0\,\tilde{\epsilon}_0) \neq 0$, then we can recast \eqref{EQ:1/2_BPS_equation_on_S2} as
\begin{align}
\td{}{t}\log\l(\f{2r\epsilon_0^{-} + \mr{i}\epsilon_0^{+}z}{2r\tilde{\epsilon}_0^{-} + \mr{i}\tilde{\epsilon}_0^{+}z} \f{2r\tilde{\epsilon}_0^{+} + \mr{i}\tilde{\epsilon}_0^{-}\bar{z}}{2r\epsilon_0^{+} + \mr{i}\epsilon_0^{-}\bar{z}}\r) = 0 \;,
\end{align}
so that the differential equation reduces to a rational one. In particular, given the condition on the determinant, we can always find a M\"obius transformation that brings the above to an equation of the form
\begin{align}\label{EQ:Mobius_equation}
\bar{z} = \f{az+b}{cz+d} \;.
\end{align}
When the coefficients allow for solutions of the above, the corresponding paths on the sphere are circles. The situation is analogous for $\det(\epsilon_0\,\tilde{\epsilon}_0)=0$. In that case one can recast \eqref{EQ:1/2_BPS_equation_on_S2} into
\begin{align}
\td{}{t}\l(\f{1}{\epsilon_0^+} \, \f{1}{\epsilon_0^+z-2\mr{i}r\epsilon_0^-} + \f{1}{\epsilon_0^-} \, \f{1}{\epsilon_0^- \bar{z} - 2\mr{i}r\epsilon_0^+}\r) = 0 \;,
\end{align}
which, again, reduces to a rational equation of the form \eqref{EQ:Mobius_equation}. Here we are assuming that certain components of $\epsilon_0$ and $\tilde{\epsilon}_0$ are non-vanishing; other corner cases do not introduce solutions which are qualitatively different from the one discussed above.

So far we have concluded that \eqref{EQ:1/2_BPS_equation_on_S2} admits solutions only for certain choices of $\epsilon_0$ and $\tilde{\epsilon}_0$, and these solutions can only be circles. Now, instead of deriving the explicit form of a generic solution of \eqref{EQ:1/2_BPS_equation_on_S2}, without loss of generality we will focus only on those circles which are centered in $z=0$, i.e.\ latitudes. Any other solution can be mapped into this restricted class by a suitable $\mr{SU}(2)$ transformation.

When substituting the ansatz $z(t) = \rho\,e^{i\varphi(t)}$, \eqref{EQ:1/2_BPS_equation_on_S2} is satisfied for either
\begin{align}
\epsilon_0^-=\tilde{\epsilon}_0^+&=0 & \t{or}& & \epsilon_0^+=\tilde{\epsilon}_0^-&=0 \;.
\end{align}
In both cases one finds
\begin{align}
\xi^\ms{a} = \tilde{\epsilon}\gamma^{\ms{a}}\epsilon \propto \begin{pmatrix} 0 \\ \sin\theta \end{pmatrix} \;,
\end{align}
i.e.\ that the Killing vector defined by the two spinors generate the $\mr{U}(1)$ isometry whose orbits on the sphere are the paths of the $\f{1}{2}$-BPS Wilson line considered. At the same time, the fixed points of the action of such isometry are the north and the south poles, where $f$ and $\tilde{f}$ have their zeros and their singularities.

As anticipated, these conclusions are fully general. In fact, one can proceed the other way around and consider any $\f{1}{2}$-BPS Wilson line on $S^2$. This will run along the action of some $\mr{U}(1)$ isometry induced by the supercharges annihilating the Wilson line. The fixed points of the isometry will be precisely the antipodal points where $f$ and $\tilde{f}$ become singular. One can then identify the two points with north and south pole and choose appropriate spherical coordinates in terms of which the $\f{1}{2}$-BPS paths have $\theta$ fixed. For any value of $\theta$, one finds that there are actually two Wilson lines annihilated by two different pair of supercharges $\v{Q} = \epsilon Q$ and $\tilde{\v{Q}} = \tilde{\epsilon} \tilde{Q}$. The first one takes the form
\begin{align}
W_{\mr{a}} &= \Pexp \int_{\varphi_0}^{\varphi_1} \l[\mr{i}A_\varphi - r\l(\cos^2\!\tf{\theta}{2}\,\sigma + \sin^2\!\tf{\theta}{2}\,\tilde{\sigma}\r)\r] \d\varphi \;,
\end{align}
with
\begin{align}\label{EQ:killing_spinors_a}
\epsilon_{\mr{a}} &= \epsilon_0^+ \, e^{-i\varphi/2} \begin{pmatrix}\mr{i}\sin\tf{\theta}{2}\\\cos\tf{\theta}{2}\end{pmatrix} \;, & \tilde{\epsilon}_{\mr{a}} &= \tilde{\epsilon}_0^- \, e^{+i\varphi/2} \begin{pmatrix}\cos\tf{\theta}{2}\\\mr{i}\sin\tf{\theta}{2}\end{pmatrix}  \;,
\end{align}
while the second reads
\begin{align}
W_{\mr{b}} &= \Pexp \int_{\varphi_0}^{\varphi_1} \l[\mr{i}A_\varphi + r\l(\sin^2\!\tf{\theta}{2}\,\sigma + \cos^2\!\tf{\theta}{2}\,\tilde{\sigma}\r)\r] \d\varphi \;,
\end{align}
with
\begin{align}\label{EQ:killing_spinors_b}
\epsilon_{\mr{b}} &= \epsilon_0^- \, e^{+i\varphi/2} \begin{pmatrix}\cos\tf{\theta}{2}\\\mr{i}\sin\tf{\theta}{2}\end{pmatrix}  \;, & \tilde{\epsilon}_{\mr{b}} &= \tilde{\epsilon}_0^+ \, e^{-i\varphi/2} \begin{pmatrix}\mr{i}\sin\tf{\theta}{2}\\\cos\tf{\theta}{2}\end{pmatrix} \;.
\end{align}

Notice how, since at the poles either $f$ or $\tilde{f}$ are singular, the integrals do not vanish in the limits $\theta\to0$ and $\theta\to\pi$. In these limits the Wilson lines reduce to the local operators
\begin{align}\label{EQ:local_operators}
\lim_{\theta\to0}W_{\mr{a}} &= e^{r(\varphi_0-\varphi_1)\sigma|_{\theta=0}} \;, & \lim_{\theta\to\pi}W_{\mr{a}} &= e^{r(\varphi_0-\varphi_1)\tilde{\sigma}|_{\theta=\pi}} \;, \cr
\lim_{\theta\to0}W_{\mr{b}} &= e^{r(\varphi_1-\varphi_0)\tilde{\sigma}|_{\theta=0}} \;, & \lim_{\theta\to\pi}W_{\mr{b}} &= e^{r(\varphi_1-\varphi_0)\sigma|_{\theta=\pi}} \;.
\end{align}
These limits will be important for our analysis in the next sections.

\subsection{The squashed sphere}\label{SEC:squashed_sphere}
The round sphere can be seen as a special case of a more general supersymmetric background. This is the squashed sphere, whose geometry is given by
\begin{align}
e^{\ms{1}} &= r \, \ell(\theta) \, \d\theta\;, \cr
e^{\ms{2}} &= r \sin\theta \, \d\varphi \;,
\end{align}
for some smooth $\ell(\theta)>0$ with $\ell(0) = \ell(\pi) = 1$.

This supersymmetric background, studied in \cite{Gomis:2012wy}, takes the form
\begin{align}
    B &= -\f{1}{2}\l(1-\f{1}{\ell(\theta)}\r)\,\d\varphi \;, & H = \tilde{H} = -\f{\mr{i}}{r \ell(\theta)} \;.
\end{align}
A generic $\ell(\theta)$ preserves only a $\mr{U}(1)$ isometry group and two supercharges associated with the Killing spinors in \eqref{EQ:killing_spinors_a}, which in turn, generate the residual isometry. Inverting the sign of $B$, leads to a different background in which the preserved supercharges are the ones in \eqref{EQ:killing_spinors_b}. Here the only $\tf{1}{2}$-BPS Wilson lines are the ones running along the action of the preserved $\mr{U}(1)$, and are of the form we denoted with $W_{\mr{a}}$ (or $W_{\mr{b}}$, for a different sign in $B$).

\section{Wilson loops}\label{SEC:WL}
In this section we will focus on Wilson loops, i.e.\ we will consider the gauge-invariant observables obtained by taking the trace, in some representation $\mc{R}$ of $\v{G}$, of a Wilson line defined on a closed path $\Gamma$,
\begin{align}
L_{\mc{R}}(\Gamma) = \tr_{\mc{R}} \Pexp \oint_{\Gamma} \mr{i}\mc{A} \;.
\end{align}
For simplicity, we will consider only non-self-intersecting Wilson loops. We will also mainly focus on Wilson loops annihilated by $\v{Q}$, although what we will say will extend straightforwardly to those annihilated by $\tilde{\v{Q}}$.

The fact that the field strength $\mc{F}$ associated with the supersymmetric connection $\mc{A}$ is $\v{Q}$-exact has the important consequence that a Wilson loop is invariant under a smooth deformation of $\Gamma$. It is possible to show, in fact, that two Wilson loops whose paths are homotopically equivalent are $\v{Q}$-cohomologous. We will prove this by showing that their difference is a $\v{Q}$-exact quantity.

We start by considering the Wilson line
\begin{align}
    W(s;t_0,t) = \Pexp \int_{t_0}^t\d t' \; \mr{i}\mc{A}_t(s,t') \;.
\end{align}
defined on a homotopy of paths $\Gamma(s;t)$, where $s$ and $t$ parametrise respectively the homotopy and the path. We also introduce the components
\begin{align}
    \mc{A}_t &= \mc{A}^\ms{a} \, \partial_tx^\ms{a} \;,
    &
    \mc{A}_s &= \mc{A}^\ms{a} \, \partial_sx^{\ms{a}} \;,
    &
    \mc{F}_{st} &= \mc{F}^{\ms{ab}} \, \partial_tx^{\ms{a}} \, \partial_sx^{\ms{b}} \;,
\end{align}
Using the variation formula
\begin{align}
    \partial_s W(s;0,1) = \mr{i}\int_0^1\d t' \; W(s;t',1) \; \partial_s\mc{A}_t(s;t') \; W(s;0,t') \;,
\end{align}
it is possible to show \cite{Menotti:1994ab} that
\begin{align}\label{EQ:cov_der_WL}
    \partial_sW(s;0,1) = \;&\mr{i}[\mc{A}_s(s;1) \; W(s;0,1) - W(s;0,1) \; \mc{A}_s(s;0)] \cr
    &+ \mr{i}\int_{0}^1\d t' \; W(s;t',1) \; \mc{F}_{st}(s;t') \; W(s;0,t')
\end{align}
Notice that if we are dealing with a closed loop, i.e.\ if $\Gamma(s;0)=\Gamma(s;1)$, the first term in the right hand side of \eqref{EQ:cov_der_WL} becomes a commutator. Therefore
\begin{align}
    \partial_s L_{\mc{R}} = \mr{i}\tr\int_{0}^1\!\!\d t' \, W(s;t',1) \, \mc{F}_{st}(s;t') \, W(s;0,t')
\end{align}
It follows from \eqref{EQ:F_exact} that
\begin{multline}\label{EQ:varW_exact}
    \partial_s L_{\mc{R}, \epsilon} = \v{Q} \tr_{\mc{R}} \int_0^1 \d t' \, \varepsilon^{\ms{ab}}\, \partial_t x^{\ms{a}}(t')\, \partial_s x^{\ms{b}}(t') \, \\ \times W_{\epsilon}(s;t',1) \l[-\f{\epsilon\lambda}{2\epsilon^+ \epsilon^-}\r] W_{\epsilon}(s;0,t') \;.
\end{multline}
Likewise, the variation of the Wilson loop annihilated by $\tilde{\v{Q}}$ is $\tilde{\v{Q}}$-exact and reads
\begin{multline}
    \partial_s L_{\mc{R}, \tilde{\epsilon}} = \tilde{\v{Q}} \tr \int_0^1 \d t' \, \varepsilon^{\ms{ab}}\, \partial_t x^{\ms{a}}(t')\, \partial_s x^{\ms{b}}(t') \, \\ \times W_{\tilde{\epsilon}}(s;t',1) \bigg[{+}\f{\tilde{\epsilon}\tilde{\lambda}}{2\epsilon^+ \epsilon^-} \bigg] W_{\tilde{\epsilon}}(s;0,t') \;.
\end{multline}

For a finite homotopy one can integrate both sides of \eqref{EQ:varW_exact} and show that the difference between two homotopic Wilson loops is indeed $\v{Q}$-exact. Crucially, in extending our argument to a finite deformation of $\Gamma$, one should be careful to avoid singularities of \eqref{EQ:varW_exact}, at which our argument breaks down. These singularities come from the zeros of $\epsilon^+$ and $\epsilon^-$, which are also the sources of singularities for $f_\epsilon$ and $\tilde{f}_{\epsilon}$. In general, one should consider not just the homotopy of $\mh{M}$ but rather the homotopy of $\mh{M}$ with punctures corresponding to the points in which the components of $\epsilon$ vanish.

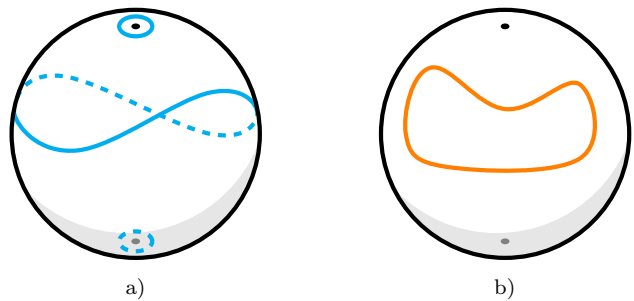
\begin{figure}[htb]
\subfloat[\label{FIG:spheres_a}]{
    \begin{tikzpicture}[ultra thick,scale=1.1]
    \path[use as bounding box] (-1.6,-1.6) rectangle (1.6,1.6);
    \colorlet{wilsonline}{cyan}
    \colorlet{shadow}{black!10}
    \begin{scope}
    \clip (-1.5,-1.5) rectangle (1.5,0.0);
    \fill [shadow] (0,0) circle[radius=1.5];
    \fill [white] (-0.1,0.5) circle[radius=1.7];
    \end{scope}
    \begin{scope}
    \clip (-1.47,0.37) -- (-1.47,1.0) -- (1.47,0.25) -- (1.47,-0.5) -- cycle;
    \draw [wilsonline, dashed] plot [smooth, tension=0.7] coordinates {(0.9,0.5) (1.47,0.25) (0.9,0.0) (-0.9,0.7) (-1.45,0.25) (-0.7,-0.2)};
    \end{scope}
    \begin{scope}
    \clip (-1.47,0.28) -- (-1.47,-0.5) -- (1.47,0.25) -- (1.47,1.0) -- cycle;
    \draw [wilsonline] plot [smooth, tension=0.7] coordinates {(-0.9,0.7) (-1.47,0.25) (-0.7,-0.2) (0.9,0.5) (1.47,0.25) (0.9,0.0)};
    \end{scope}
    \draw (0,0) circle (1.5cm);
    \draw [wilsonline](0,1.3) ellipse (0.2cm and 0.12cm);
    \fill [black](0,1.3) ellipse (1.5pt and 1pt);
    \draw [wilsonline, dashed](0,-1.3) ellipse (0.2cm and 0.12cm);
    \fill [black!50](0,-1.3) ellipse (1.5pt and 1pt);
    \end{tikzpicture}
}
\hfill
\subfloat[\label{FIG:spheres_b}]{
    \begin{tikzpicture}[ultra thick,scale=1.1]
    \path[use as bounding box] (-1.6,-1.6) rectangle (1.6,1.6);
    \colorlet{wilsonline}{orange}
    \colorlet{shadow}{black!10}
    \begin{scope}
    \clip (-1.5,-1.5) rectangle (1.5,0.0);
    \fill [shadow] (0,0) circle[radius=1.5];
    \fill [white] (-0.1,0.5) circle[radius=1.7];
    \end{scope}
    \draw [wilsonline] plot [smooth cycle, tension=0.7] coordinates {(0.9,-0.3) (0.9,0.6) (0.0,0.3) (-0.9,0.8) (-1.2,0.0) (-0.7,-0.4)};
    \draw (0,0) circle (1.5cm);
    \fill [black](0,1.3) ellipse (1.5pt and 1pt);
    \fill [black!50](0,-1.3) ellipse (1.5pt and 1pt);
    \end{tikzpicture}
}
\caption{In (a) the Wilson loop can be mapped to local operators by shrinking its contour around the poles. In (b) the 0-homotopic Wilson loop can be mapped to $\tr \v{1}$.}\label{FIG:spheres}
\end{figure}

This property is already important at genus zero. In fact, because of homotopic invariance, one might be tempted to conclude that, at genus zero, all Wilson loops are necessarily trivial. However, while this is true for the Euclidean plane, this turns out not to be the case for the squashed sphere, despite the fact that $\pi_1(S^2)=0$. This is precisely because, as noted in Section~\ref{SEC:sphere}, for our choice of supercharges, the Killing spinors are singular at north and south pole. When these two points are removed, different contours, which are homotopically inequivalent, lead to inequivalent Wilson loops as depicted in FIG.~\ref{FIG:spheres}.

\section{Localisation at genus zero}
\label{SEC:localisation}
The aim of this section is to find an exact expression for supersymmetric Wilson loops on the squashed sphere, and their correlators. Exact results for $\mc{N}=(2,2)$ gauge theories have been derived in recent years through supersymmetric localisation \cite{Benini:2012ui, Doroud:2012xw, Gomis:2012wy}. All these results are obtained with a choice of localising supercharge $\mc{Q}$ which is a combination of $\v{Q}$ and $\tilde{\v{Q}}$. Since a generic Wilson loop defined as in either \eqref{EQ:W_L} or \eqref{EQ:W_R} is only annihilated by one chiral supercharge, one cannot directly apply the recipe of \cite{Benini:2012ui, Doroud:2012xw, Gomis:2012wy} to compute the expectation value of such a Wilson loop. However, the conclusions of the previous section come to rescue, as one can use the invariance under homotopy to map a generic Wilson loop to a local $\f{1}{2}$-BPS operator. If we consider, for instance, a Wilson loop of the kind depicted in FIG.~\ref{FIG:spheres_a}, one can compute its expectation value by simply considering the associated local operator inserted at the pole obtained by shrinking the contour as in \eqref{EQ:local_operators}. Specifically, for a Wilson loop annihilated by $\v{Q} = \epsilon_{\mr{a}} Q$ and wrapping the north pole anticlockwise one finds
\begin{align}\label{EQ:operator_insertion}
\langle L_{\mc{R}} \rangle = \langle \tr_{\mc{R}} e^{-2\pi r\sigma} \rangle \;.
\end{align}
A change in orientation will result in a simple sign flip in the exponent.
In a similar fashion, one can recast a correlator of $n$ Wilson loops annihilated by the same $\v{Q}$ as
\begin{multline}\label{EQ:correlator}
\big\langle L_{\mc{R}_1}(\Gamma_1) \ldots L_{\mc{R}_n}(\Gamma_n) \big\rangle = \prod_{\Gamma_i \in [0]} \dim\mc{R}_i  \\
\times \bigg\langle \prod_{\Gamma_j \in [1]}\tr_{\mc{R}_j} e^{-2\pi r\sigma|_{\theta=0}} \prod_{\Gamma_k \in [-1]}\tr_{\mc{R}_k} e^{+2\pi r\sigma|_{\theta=0}} \bigg\rangle \;,
\end{multline}
where $[n]$ is the homotopy class of paths that wind $n$ times around the north pole.

What is crucial for the success of this approach is that these local operators are annihilated by two supercharges, as noted in Section~\ref{SEC:sphere}. In particular, they are annihilated by $\mc{Q}$, and as such, are amenable to localisation.
Here we get to the final result by effectively using two cohomological arguments. The first, with respect to the supercharge $\v{Q}$, uses the invariance under a variation of the homotopy parameter $s$ to map any Wilson loop to a $\f{1}{2}$-BPS local operator. The second, with respect to the supercharge $\mc{Q}$, uses the invariance under a smooth variation of the gauge coupling (and possibly other parameters) to reduce the path integral to a Coulomb-branch matrix model as in \cite{Benini:2012ui, Doroud:2012xw, Gomis:2012wy}. We will now briefly summarise the setup for the computation.

Let us consider a theory of a vector multiplet and chiral multiplets, with gauge group $\v{G}$. The Lagrangian for the vector multiplet is given by
\begin{align}
    \mh{L}_{\mr{vec}} &= \tr\Bigg[ \f{1}{2}\l(*F - \mr{i}\f{\sigma - \tilde{\sigma}}{2r\ell(\theta)} \r)^2 + \f{1}{2}\l(D + \f{\sigma + \tilde{\sigma}}{2r\ell(\theta)} \r)^2 \cr
    &\quad + \f{1}{2}\mr{D}^{\ms{a}}\tilde{\sigma}\mr{D}^{\ms{a}}\sigma -\f{1}{8}[\sigma, \tilde{\sigma}]^2 - \f{\mr{i}}{2}\tilde{\lambda}\,\gamma^{\ms{a}}\mr{D}^{\ms{a}}\lambda \cr
    &\quad - \f{\mr{i}}{2}\tilde{\lambda}\mr{P}_+[\sigma,\lambda] - \f{\mr{i}}{2}\tilde{\lambda}\mr{P}_-[\tilde{\sigma},\lambda]\Bigg] \;.
\end{align}
For every generator in the centre of $\mf{g}$ we can add a topological and a Fayet--Iliopoulus term, namely
\begin{align}
    \mh{L}_{\mr{FI}} = \mr{i}\f{\vartheta}{2\pi}\tr *F - \mr{i}\xi \tr D \;.
\end{align}
In the following we will rely on the presece of such a term.
The chiral multiplet has components $(\phi,\psi,F)$, while the anti-chiral has components $(\tilde\phi,\tilde\psi,\tilde{F})$. The Lagrangian for the matter content is
\begin{align}
    \mh{L}_{\mr{mat}} &= \tilde\phi\l[-\mr{D}^{\ms{a}}\mr{D}^{\ms{a}} + \tilde{\sigma}\sigma + \f{\mr{i}q(\sigma+\tilde\sigma)}{2r\ell(\theta)} + \f{q(2-q)}{4r^2\ell^2(\theta)}\r]\phi \cr
    &\quad +\mr{i}\tilde\psi\l[-\gamma^{\ms{a}}\mr{D}^{\ms{a}} + \sigma\mr{P}_- + \tilde\sigma\mr{P}_+ +\f{\mr{i}q}{2r\ell(\theta)} \r]\psi \cr
    &\quad +\mr{i}\tilde{\psi}\lambda\phi - \mr{i}\tilde{\phi}\psi \;.
\end{align}
This can be complemented by the introduction of superpotential interactions $\mh{L}_{\mr{pot}}$. The choice of the superpotential will determine the R-charge assignments for the matter fields. Finally, if the theory has some flavour group $\v{G}_{\mr{F}}$, we can gauge it by introducing a non-dynamical vector multiplet and turning on background values for its bosonic component fields along the Cartan \cite{Benini:2012ui, Doroud:2012xw, Park:2016dpb}.
These will introduce twisted masses $\tau_i/r$ and background monopole charges $\mf{n}_i$, where $i=1, \dots, \rk\v{G}_{\mr{F}}$. 

It is possible to localise this theory \cite{Benini:2012ui, Doroud:2012xw} with respect to a charge
\begin{align}
\mc{Q} = \v{Q}|_{\epsilon_0^+=1} + \tilde{\v{Q}}|_{\tilde{\epsilon}_0^-=-\mr{i}} \;.
\end{align}
In Coulomb-branch localisation one uses the fact that $\mh{L}_{\mr{vec}}$, $\mh{L}_{\mr{mat}}$ and $\mh{L}_{\mr{pot}}$ are all \mbox{$\mc{Q}$-exact}. This implies that the expectation value of $\mc{Q}$-closed observables will not depend on the couplings appearing in the action for the vector and the chiral multiplet, and on parameters in the superpotential. The result will depend, instead, on the parameters $\vartheta$ and $\xi$ in the Fayet--Iliopoulus action and on the background flavour gauge multiplet through $\tau_i$ and $\mf{n_i}$.

The BPS locus is spanned by the field configurations
\begin{align}\label{EQ:BPS_locus}
A &= \f{\mf{m}}{2}(\kappa - \cos\theta) \, \d\varphi \;, \cr
\sigma &= \f{2y-\mr{i}\mf{m}}{2r} \;, \cr
\tilde{\sigma} &= \f{2y+\mr{i}\mf{m}}{2r} \;, \cr
D &= -\f{y}{r^2} \;,
\end{align}
aligned with the Cartan. Here $y\in\mb{R}^{\rk\v{G}}$, $\mf{m}\in\mb{Z}^{\rk\v{G}}$ are monopole charges and $\kappa$ is chosen to be $+1$ in a coordinate patch that covers the north pole, and $-1$ in one covering the south pole. The one-loop determinant associated with the vector multiplet reads
\begin{align}
    Z_{\rm{vec}}(\mf{m},y) = \prod_{\alpha\in\v{G}}\l(\f{\alpha(\mf{m})^2}{4} + \alpha(y)^2 \r) \;,
\end{align}
where the product is over the roots $\alpha$ of $\v{G}$. Then consider a chiral multiplet with R-charge $q$ in a representation $\mc{R}_{\mr{mat}}$ of $\v{G}$ and in a representation of $\v{G}_{\mr{F}}$ with charges $h^i$. Its one-loop determinant is given by
\begin{multline}\label{EQ:Z_mat}
    Z_{\rm{mat}}(\mf{m},y;\mf{n},\tau) \\ = \!\!\prod_{\rho\in\mc{R}_{\mr{mat}}}\!\!\f{\Gamma\l(\tf{1}{2}q - \mr{i}\rho(y) - \mr{i}h^i\tau_i - \tf{1}{2}\rho(\mf{m}) - \tf{1}{2}h^i\mf{n}_i \r)}{\Gamma\l(1 - \tf{1}{2}q + \mr{i}\rho(y) + \mr{i}h^i\tau_i -\tf{1}{2}\rho(\mf{m}) - \tf{1}{2}h^i\mf{n}_i \r)} \;,
\end{multline}
where the product is over the weight $\rho$ of $\mc{R}$. The contribution associated with the classical action comes from $\mh{L}_{\mr{FI}}$. This, evaluated on the locus, gives
\begin{align}
    Z_{\rm{cl}}(\mf{m},y) = e^{-4\pi\mr{i}\xi \tr y-\mr{i}\vartheta \tr\mf{m}} \;,
\end{align}
When putting things together, Coulomb-branch localisation gives rise to the matrix model
\begin{align}\label{EQ:Coulomb_branch_formula}
Z = \!\! \sum_{\mf{m}\in\mb{Z}^{\rk\v{G}}}\!\int \!\prod_{r=1}^{\rk\v{G}}\!\f{\d y_r}{2\pi} \, Z_{\rm{cl}}(\mf{m},y) \, Z_{\mr{vec}}(\mf{m},y) \, Z_{\mr{mat}}(\mf{m},y;\mf{n},\tau) .
\end{align}

One can compute the expectation value of a $\mc{Q}$-closed operator $\mc{O}$ through the above matrix model simply by considering the insertion of $\mc{O}$ evaluated on the BPS locus \eqref{EQ:BPS_locus}. The expectation value in \eqref{EQ:operator_insertion}, in particular, corresponds to the insertion of
\begin{align}\label{EQ:MM_operator_insertion}
\tr_{\mc{R}} e^{-2\pi r \sigma}|_{\t{BPS locus}} = \sum_{\Lambda\in\mc{R}} e^{-2\pi \Lambda(y) + i \pi \Lambda(\mf{m})} \;.
\end{align}

In the following we will consider theories with unitary gauge groups.

\subsection{\texorpdfstring{$\mr{U}(1)$}{U(1)} gauge theory with matter}
We will start by considering a theory of gauge group $\mr{U}(1)$ with the $N_\mr{f}$ chiral multiplets of charge $+1$ and twisted masses $\tau_f$, $N_\mr{a}$ chiral multiplets of charge $-1$ and twisted masses $\tilde{\tau}_a$, and vanishing background flavour fluxes. The matrix model resulting from Coulomb-branch localisation reads
\begin{multline}\label{EQ:U(1)_matrix_model}
    Z_{\mr{U}(1)}(\xi,\vartheta; \tau, \tilde{\tau}) = \sum_{\mf{m}\in\mb{Z}}\int\f{\d y}{2\pi} \; e^{-4\pi\mr{i}\xi y-\mr{i}\mf{m}\vartheta} \\
    \times \prod_{f=1}^{N_\mr{f}}\f{\Gamma(-\mr{i} y-\mr{i}\tau_f-\mf{m}/2)}{\Gamma(1+\mr{i} y+\mr{i}\tau_f-\mf{m}/2)} \prod_{a=1}^{N_\mr{a}}\f{\Gamma(\mr{i} y-\mr{i}\tilde{\tau}_f+\mf{m}/2)}{\Gamma(1-\mr{i} y+\mr{i}\tilde{\tau}_a+\mf{m}/2)} \;.
\end{multline}
The R-charge contributions can reabsorbed by giving a imaginary part to the twisted masses, as can be seen in \eqref{EQ:Z_mat}.

Without loss of generality we will assume that $N_{\mr{f}} > N_{\mr{a}}$, or that $N_{\mr{f}} = N_{\mr{a}}$ and $\xi>0$. The other cases can obtained by charge conjugation.

The issue about the convergence of the above matrix model with the insertion of the Wilson loop operator can be simply addressed by using the fact that the partition function is analytic in $\xi$ and $\vartheta$ (see \cite{Benini:2012ui}).
In the abelian case, then, the insertion of the local operator \eqref{EQ:MM_operator_insertion} corresponds to a shift in the parameters $\vartheta$ and $\mf{m}$. In particular
\begin{align}\label{EQ:loc_U1}
    \l\langle e^{-2\pi r\sigma}\r\rangle_{\mr{U}(1)} = \f{Z_{\mr{U}(1)}(\xi-\Lambda\tf{\mr{i}}{2}, \vartheta-\Lambda\pi; \tau,\tilde{\tau})}{Z_{\mr{U}(1)}(\xi,\vartheta;\tau,\tilde{\tau})} \;,
\end{align}
where $\Lambda$ is the weight of $\mc{R}$, i.e.\ an integer labelling the Wilson-loop representation.
The integral representation in \eqref{EQ:U(1)_matrix_model} can be recast into the ``Higgs-branch'' formula \cite{Benini:2012ui, Doroud:2012xw}

\begin{multline}
    Z_{\mr{U}(1)}(\xi,\vartheta;\tau,\tilde{\tau}) \\ = \sum_{l=1}^{N_\mr{f}}e^{4\pi\mr{i}\xi\tau_l} \, Z^{(l)}_{\text{1-loop}}(\tau,\tilde{\tau}) \, Z^{(l)}_\mr{v}(z;\tau,\tilde{\tau}) \, Z^{(l)}_\mr{av}(\bar{z};\tau,\tilde{\tau}) \;,
\end{multline}
where we set $z=e^{-2\pi\xi-\mr{i}\vartheta}$ and $\bar{z}=e^{-2\pi\xi+\mr{i}\vartheta}$, and the functional determinants can be written in terms of hypergeometric functions as
\begin{align}
    Z^{(l)}_{\text{1-loop}}(\tau,\tilde{\tau}) &= \prod_{\substack{f=1 \\ f \neq l}}^{N_\mr{f}} \f{\Gamma(-\mr{i}M_f^l)}{\Gamma(1+\mr{i}M_f^l)}\prod_{a=1}^{N_\mr{a}}\f{\Gamma(-\mr{i}\tilde{M}_a^l)}{\Gamma(1+\mr{i}\tilde{M}_a^l)} \;, \\
    Z^{(l)}_\mr{v}(z;\tau,\tilde{\tau}) &= \hG{N_\mr{a}}{N_\mr{f}-1}{\{-\mr{i}\tilde{M}_a^l \}_{a=1}^{N_\mr{a}}}{\{1+\mr{i}M_f^l \}_{f=1, f\neq l}^{N_\mr{f}}}{(-1)^{N_\mr{f}}z} , \\
    Z^{(l)}_\mr{av}(\bar{z};\tau,\tilde{\tau}) &= \hG{N_\mr{a}}{N_\mr{f}-1}{\{-\mr{i}\tilde{M}_a^l\}_{a=1}^{N_\mr{a}}}{\{1+\mr{i}M_f^l\}_{f=1, f \neq l}^{N_\mr{f}}}{(-1)^{N_\mr{a}}\bar{z}} ,
\end{align}
where $M_f^l=\tau_f-\tau_l$ and $\tilde{M}_a^l=\tilde{\tau}_f+\tilde{\tau}_l$.
\vspace{1.em}

\subsection{\texorpdfstring{$\mr{U}(N)$}{U(N)} gauge theory with matter}
We now want to generalise the result above to the case of a $\mr{U}(N)$ gauge group with matter in the fundamental and anti-fundamental representation. To avoid supersymmetry breaking we consider theories with $N \leq N_{\mr{f}}$. Coulomb branch localisation leads to the matrix model
\begin{widetext}
\begin{align}
    Z_{\mr{U}(N)}(\xi,\vartheta;\tau,\tilde{\tau}) &= \f{1}{N!}\sum_{\mf{m}\in\mb{Z}^N}\int \prod_{r=1}^N\f{\d y_r}{2\pi} \; e^{-4\pi\mr{i}\xi y_r-\mr{i}\mf{m}_r\vartheta} \prod_{1\leq t < s \leq N}\!\l(\tf{1}{4}(\mf{m}_t - \mf{m}_s)^2 + (y_t - y_s)^2 \r) \cr &\hspace*{1cm} \times\prod_{r=1}^N\l[\prod_{f=1}^{N_\mr{f}}\f{\Gamma(-\mr{i}y_r-\mr{i}\tau_f-\mf{m}_r/2)}{\Gamma(1+\mr{i}y_r+\mr{i}\tau_f-\mf{m}_r/2)}\,\prod_{a=1}^{N_\mr{a}}\f{\Gamma(\mr{i}y_r-\mr{i}\tilde{\tau}_a+\mf{m}_r/2)}{\Gamma(1-\mr{i}y_r+\mr{i}\tilde{\tau}_a+\mf{m}_r/2)} \r] \;.
\end{align}
\end{widetext}
We notice that it is possible to obtain $Z_{\mr{U}(N)}$ by acting with a differential operator $\Delta$ on $N$ copies of $Z_{\mr{U}(1)}$. Namely,
\begin{align}
    Z_{\mr{U}(N)}(\xi,\vartheta;\tau,\tilde{\tau}) = \f{1}{N!} \; \Delta\prod_{r=1}^N Z_{\mr{U}(1)}(\xi_r,\vartheta_r;\tau,\tilde{\tau})\Big|_{\subalign{\xi_r &= \xi \\ \vartheta_r &= \vartheta}} \;,
\end{align}
where
\begin{align}
    \Delta = \!\!\prod_{1 \leq t < s \leq N} \l[-\f{1}{4}\l(\pd{}{\vartheta_r} - \pd{}{\vartheta_s} \r) - \f{1}{16\pi} \l(\pd{}{\xi_r} - \pd{}{\xi_s} \r) \r] .
\end{align}
Similarly, the insertion of \eqref{EQ:MM_operator_insertion} is obtained by evaluating
\begin{multline}
    \l\langle \tr_{\mc{R}} e^{-2\pi r\sigma} \r\rangle_{\mr{U}(N)} = \l[Z_{\mr{U}(N)}(\xi,\vartheta;\tau,\tilde{\tau})\r]^{-1} \\ \times \sum_{\Lambda\in\mc{R}} \Delta \prod_{r=1}^N Z_{\mr{U}(1)}(\xi_r,\vartheta_r;\tau,\tilde{\tau})\bigg|_{\subalign{\xi_r&=\xi-\Lambda_r\mr{i}/2 \\ \vartheta_r&=\vartheta-\Lambda_r\pi}} \;. \\
\end{multline}
Following \cite{Benini:2012ui}, we introduce new coordinates
\begin{align}
    w_r=-\log z_r = 2\pi\xi_r + \mr{i}\vartheta_r
\end{align}
and their complex conjugates $\bar{w}_r$. For the partition function, we have
\begin{multline}\label{EQ:Higgs_branch_formula}
    Z_{\mr{U}(N)}(\xi,\vartheta;\tau,\tilde{\tau}) = \sum_{l \in \ms{C}(N_\mr{f},N)}e^{4\pi\mr{i}\xi\sum_r\tau_{l_r}} \\ \times \mc{Z}^{(l)}_{\text{1-loop}}(\tau,\tilde{\tau}) \, \mc{Z}^{(l)}_\mr{v}(w;\tau,\tilde{\tau}) \, \mc{Z}^{(l)}_{\mr{av}}(\bar{w};\tau,\tilde{\tau}) \;,
\end{multline}
where $l=(l_1,\dots,l_N)$ is a combination $\ms{C}$ of $N$ elements out of $N_\mr{f}$, and the functional determinants are
\begin{widetext}
\begin{align}
    \mc{Z}^{(l)}_{\text{1-loop}}(\tau,\tilde{\tau}) &= \Bigg[\prod_{1 \leq t < s \leq N}\l(M_{l_s}^{l_t}\r)^2 \Bigg] \Bigg[\prod_{r=1}^N Z^{(l_r)}_{\text{1-loop}} \Bigg] \;, \\
    \mc{Z}^{(l)}_\mr{v}(w;\tau,\tilde{\tau}) & = \prod_{1\leq t < s \leq N}\l(1 - \f{\partial_t - \partial_s}{\mr{i}M_{l_s}^{l_r}}\r) \; \l[\prod_{r=1}^N Z^{(l_r)}_\mr{v}(e^{-w_r};\tau,\tilde{\tau})\r]\Bigg|_{w_r=w} \;, \\
    \mc{Z}^{(l)}_\mr{av}(\bar{w};\tau,\tilde{\tau}) &= \prod_{1\leq t < s \leq N}\l(1 - \f{\bar{\partial}_t-\bar{\partial}_s}{\mr{i}M_{l_s}^{l_t}} \r) \; \l[\prod_{r=1}^N Z^{(l_r)}_\mr{av}(e^{-\bar{w}_r};\tau,\tilde{\tau})\r]\Bigg|_{\bar{w}_r=\bar{w}} \;;
\end{align}
here we denoted by $\partial_r = \partial/\partial w_r$ and by $\bar{\partial}_r = \partial/\partial \bar{w}_r$. Notice that the vortex and the anti-vortex functional determinant can be written in a nicer form as
\begin{align}
    \mc{Z}^{(l)}_\mr{v}(w;\tau,\tilde{\tau}) &= \sum_{k\in\mb{N}^N}(-1)^{(N-1)|k|}e^{-w|k|}\prod_{r\in l}\f{\prod_{a=1}^{N_\mr{a}}(-\mr{i}\tilde{M}_a^r)_{\tilde{z}_r}}{\prod_{f\in l}(-\mr{i}M_f^r-\tilde{z}_r)_{\tilde{z}_r}\prod_{f\not\in l}(-\mr{i}M_f^r-\tilde{z}_r)_{\tilde{z}_r}} \;, \\
    \mc{Z}^{(l)}_{\mr{av}}(\bar{w};\tau,\tilde{\tau}) &= \sum_{k\in\mb{N}^N}(-1)^{(N+N_\mr{a}-N_\mr{f}-1)|k|}e^{-\bar{w}|k|}\prod_{r\in l}\f{\prod_{a=1}^{N_\mr{a}}(-\mr{i}\tilde{M}_a^r)_{\tilde{z}_r}}{\prod_{f\in l}(-\mr{i}M_f^r-\tilde{z}_r)_{\tilde{z}_r}\prod_{f\not\in l}(-\mr{i}M_f^r-\tilde{z}_r)_{\tilde{z}_r}} \;,
\end{align}
where $|k|=\sum_r k_r$, and $\tilde{z}_r = z_p$ with $p$ such that $l_p=r$. The computation of the operator insertion is similar: with the new variables $w$ and $\bar{w}$ we have
\begin{align}\label{EQ:W_UN}
    \l\langle \tr_{\mc{R}} e^{-2\pi r\sigma}\r\rangle_{\mr{U}(N)} &= \l[Z_{\mr{U}(N)}(\xi,\vartheta;\tau,\tilde{\tau})\r]^{-1}\!\!\f{1}{N!}(-1)^{\tf{N(N-1)}{2}} \sum_{\Lambda\in\mc{R}} \Bigg\{\sum_{l_1=1}^{N_{\mr{f}}} \dots \sum_{l_N=1}^{N_\mr{f}} \l(\prod_{r=1}^N Z^{(l)}_{\text{1-loop}}(\tau,\tilde{\tau})\r) \cr
    & \hspace*{10mm}\times\prod_{1 \leq t < s \leq N}(\partial_t - \partial_s)\l[\prod_{r=1}^Ne^{\mr{i}\sum_rw_r\tau_{l_r}}Z^{(l_r)}_\mr{v}(e^{-w_r};\tau,\tilde{\tau}) \r]\Bigg|_{w_r=w-2\pi\mr{i}\Lambda_r} \cr
    & \hspace*{10mm}\times\prod_{1 \leq t < s \leq N}(\bar{\partial}_t - \bar{\partial}_s)\l[\prod_{r=1}^Ne^{\mr{i}\sum_r\bar{w}_r\tau_{l_r}}Z^{(l_r)}_\mr{av}(e^{-\bar{w}_r};\tau,\tilde{\tau})\r]\Bigg|_{\bar{w}_r=\bar{w}} \Bigg\} \;.
\end{align}
\end{widetext}
The last line of \eqref{EQ:W_UN} suppresses the the $l$'s containing repeated indices. As before, this gives a sum over $l\in\ms{C}(N_\mr{f},N)$. Keeping into account that
\begin{align}
    e^{\mr{i}\sum_rw_r\tau_{l_r}}\Big|_{w_r = w - 2\pi\mr{i}\Lambda_r} = e^{2\pi\Lambda(\tau_l)} \, e^{\mr{i}w\sum_r\tau_{l_r}} \;,
\end{align}
we finally arrive at the expression
\begin{align}
    \l\langle \tr_{\mc{R}} e^{-2\pi r\sigma} \r\rangle_{\mr{U}(N)} = \big\langle\kern-0.22em\big\langle \chi_{\mc{R}}(x_{l_1}, \ldots, x_{l_N}) \big\rangle\kern-0.22em\big\rangle \;,
\end{align}
where with $\langle\kern-0.17em\langle \mc{O} \rangle\kern-0.17em\rangle $ we denote the insertion of $\mc{O}$ in the Higgs-branch formula \eqref{EQ:Higgs_branch_formula}. Here, we have defined $x_i = e^{2\pi\tau_i}$.

We notice that an analogous conclusion could be obtained, perhaps more directly, by using Higgs branch localisation as in \cite{Benini:2012ui, Doroud:2012xw}. In this case, the localisation locus for the bottom components of the vector multiplet is given by
\begin{align}
    \sigma = \tilde{\sigma} = -\tau_l/r \;.
\end{align}
Interestingly, in order to perform Higgs branch localisation one has to assume the presence of a Fayet--Iliopoulus term, which in our case was crucial for the convergence of the matrix model.

When $N=N_{\mr{f}}$ there is only one $l$ in the sum and the result takes the simple form
\begin{align}
\l\langle \tr_{\mc{R}} e^{-2\pi r\sigma} \r\rangle_{\mr{U}(N_{\mr{f}})} = \chi_{\mc{R}}(x_1, \ldots, x_{N_{\mr{f}}})\;.
\end{align}

At this point we have everything in place to address the original aim of this section, namely to give a quantitative description of the correlator in \eqref{EQ:correlator}.
Because of the isomorphism between the representation ring and the character ring of any compact $\v{G}$, a correlator can be always recast as a single Wilson loop insertion associated with a product of representations. These are the representations (or their conjugates, according to the path orientation) of homotopically nontrivial $\Gamma$'s in the correlator. Namely,
\begin{multline}
    \big\langle L_{\mc{R}_1}(\Gamma_1) \ldots L_{\mc{R}_n}(\Gamma_n) \big\rangle_{\mr{U}(N)} = \prod_{\Gamma_i \in [0]} \dim\mc{R}_i \\ \times
    \bigg\langle\kern-0.48em\bigg\langle \chi_{\vphantom{\big|}
    \!\!\bigotimes\limits_{\Gamma_j \in [1]}\!\!\mc{R}_j \bigotimes\limits_{\Gamma_k \in [-1]}\!\!\!\!\overline{\mc{R}}_k} (x_{l_1}, \ldots, x_{l_N}) \bigg\rangle\kern-0.48em\bigg\rangle \;.
\end{multline}
Again, for $N=N_{\mr{f}}$, one simply obtains
\begin{multline}
    \big\langle L_{\mc{R}_1}(\Gamma_1) \ldots L_{\mc{R}_n}(\Gamma_n) \big\rangle_{\mr{U}(N_{\mr{f}})} \\ =
    \chi_{\vphantom{\big|}
    \!\!\bigotimes\limits_{\Gamma_j \in [1]}\!\!\mc{R}_j \bigotimes\limits_{\Gamma_k \in [-1]}\!\!\!\!\overline{\mc{R}}_k} (x_1, \ldots, x_{N_{\mr{f}}}) \prod_{\Gamma_i \in [0]} \dim\mc{R}_i \;.
\end{multline}

\section{Dualities}\label{SEC:dualities}
Two-dimensional $\mc{N}=(2,2)$ gauge theories enjoy a set of dualities \cite{Hori:2006dk, Hori:2011pd} which and are reminiscent of Seiberg duality in four dimensions \cite{Seiberg:1994pq}. For models with unitary gauge groups, the ones we are interested in, the duality, suggested by the brane construction in \cite{Hanany:1997vm}, relates a $\mr{U}(N)$ theory with $N_{\mr{f}}>N$ chiral multiplets in the fundamental representation with a $\mr{U}(N_{\mr{f}}-N)$ theory with the same matter content, under the following identification of parameters:
\begin{align}\label{EQ:duality_parameters}
    \xi^{\mr{D}} &= \xi \;, & \theta^{\mr{D}} &= \theta - N_{\mr{f}}\pi \;, & \tau^{\mr{D}} &= -\tau \;.
\end{align}

In the absence of antifundamental multiplets, the flavour symmetry group $\v{G}_{\mr{F}}$ is $\mr{SU}(N_{\mr{f}})$, which implies that $\sum_f \tau_f = 0$ and, in turn, that $\prod_f x_f=1$.

In this section we provide additional evidence for such dualities by extending their dictionary with the supersymmetric Wilson loops defined in the present work. What we find is that a single Wilson loop in a given representation of $\mr{U}(N)$ is mapped, under duality, to a linear combination of Wilson loops in different representations of $\mr{U}(N_{\mr{f}}-N)$, similarly to what was found in \cite{Kapustin:2013hpk} for theories in three dimensions.

Without loss of generality, we will write down the duality map for a single Wilson loop in an irreducible representation of $\mr{U}(N)$. An irreducible representation $\mc{R}$ is uniquely determined by its highest weight $\v{\lambda}$, so in the remainder of this section we will use the two interchangebly. We refer the reader to Appendix \ref{APP:characters}, where we provide details about $\mr{U}(N)$ characters along with relevant mathematical identities.

In the Higgs branch formula \eqref{EQ:Higgs_branch_formula}, one has to sum over combinations $l$. Every $l \in \ms{C}(N_\mr{f},N)$ has a natural dual $l^{\mr{D}} \in \ms{C}(N_\mr{f},N_\mr{f}-N)$, such that $l\cap l^{\mr{D}} = \emptyset$. Indeed, as proven in \cite{Benini:2012ui, Doroud:2012xw}, the duality is realised in \eqref{EQ:Higgs_branch_formula} at the level of individual terms in the the sum, where a term labelled by a certain $l$ is equal to the term of the dual theory labelled by the dual $l^{\mr{D}}$, with the correct identification of parameters as in \eqref{EQ:duality_parameters}.

As it turns out, this property also holds when extending the duality to Wilson loops. In fact, we can construct the dictionary for such operators by matching, term by term in the sum, both sides of the duality. Explicitly, one starts from the identity
\begin{multline}
    \chi^{\mr{U}(N)}_{\v{\lambda}}(x_{l_1}, \ldots, x_{l_N}) = \sum_{\v{\mu}} c_{\v{\mu}}(x_1, \ldots, x_{N_{\mr{f}}}) \\
    \times \chi^{\mr{U}(N_{\mr{f}}-N)}_{\v{\mu}}(x_{l^{\mr{D}}_1}, \ldots, x_{l^{\mr{D}}_{N - N_{\mr{f}}}}) \;,
\end{multline}
and finds, after applying the map \eqref{EQ:duality_examples} that prescribes $x^{\mr{D}} = x^{-1}$,
\begin{multline}
    \big\langle\kern-0.28em\big\langle \chi^{\mr{U}(N)}_{\v{\lambda}}(x_{l_1}, \ldots, x_{l_N}) \big\rangle\kern-0.28em\big\rangle = \sum_{\v{\mu}} c_{-\v{\mu}}(x^{\mr{D}}_1, \ldots, x^{\mr{D}}_{N_{\mr{f}}}) \\
    \times \big\langle\kern-0.28em\big\langle \chi^{\mr{U}(N_{\mr{f}}-N)}_{-\v{\mu}}(x^{\mr{D}}_{l^{\mr{D}}_1}, \ldots, x^{\mr{D}}_{l^{\mr{D}}_{N - N_{\mr{f}}}}) \big\rangle\kern-0.28em\big\rangle_{\mr{D}} \;.
\end{multline}
The coefficients $c_{\v{\mu}}$ are symmetric in all $x$'s. This, in particular, means that they don't carry a dependence on $l$ and, as such, are taken out of the sum over $l$ in the Higgs-branch formula.
The duality dictionary is fully specified by the explicit expression of the coefficients $c_{\v{\mu}}$.

The algorithm to determine the $c_{\v{\mu}}$'s consists of three main steps, whose technical details are given in Appendices \ref{APP:from_characters_to_power_sum}, \ref{APP:manipulating_power_sum} and \ref{APP:from_power_sum_to_characters} respectively.

In the first step one decomposes the character $\chi^{\mr{U}(N)}_{\v{\lambda}}$ as a linear combination of power sums. Particular care is needed when $\v{\lambda}$ contains some negative entries and one cannot straightforwardly apply Frobenius formula. In the second step one manipulates the power sums so that they either depend on the $x$'s of the dual combinations $l^{\mr{D}}$, or on all $x$'s. Finally, in the third step we decompose all the power sums as $\mr{U}(N_{\mr{f}})$ and $\mr{U}(N_{\mr{f}}-N)$ characters. The former will recombine to form the coefficients $c_{\v{\mu}}$.

The discussion so far has been somewhat abstract, so we now look at a concrete example for a simple but nontrivial case. We consider a Wilson loop in the adjoint representation of $\mr{U}(2)$ with $N_{\mr{f}}=3$. For ease of notation we fix $l = \{1, 2 \}$ and $l^{\mr{D}}=\{3 \}$.
\begin{widetext}
Following Appendix~\ref{APP:from_characters_to_power_sum} we find
\begin{align}
    \chi^{\mr{U}(2)}_{(1,-1)}(x_1,x_2) = -\tf{1}{2}p_{(0)}(x_1,x_2) + p_{(1,-1)}(x_1,x_2) \;.
\end{align}
Then, as in Appendix~\ref{APP:manipulating_power_sum} we express the power sum in terms of dual variables and get
\begin{align}
    \chi^{\mr{U}(2)}_{(1,-1)}(x_1,x_2) = p_{(1,-1)}(x_1,x_2,x_3) - p_{(1)}(x_1,x_2,x_3)\; p_{(-1)}(x_3) - p_{(-1)}(x_1,x_2,x_3)\; p_{(1)}(x_3) \;,
\end{align}
Finally, following Appendix~\ref{APP:from_power_sum_to_characters} we rewrite every term  using characters of $\mr{U}(3)$ and $\mr{U}(1)$, i.e.
\begin{align}
    p_{(1)}(x_1,x_2,x_3)\; p_{(-1)}(x_3) &= \chi^{\mr{U}(3)}_{(1,0,0)}(x_1,x_2,x_3)\; \chi^{\mr{U}(1)}_{(-1)}(x_3) \;, \cr
    p_{(-1)}(x_1,x_2,x_3)\; p_{(1)}(x_3) &= \chi^{\mr{U}(3)}_{(0,0,-1)}(x_1,x_2,x_3)\; \chi^{\mr{U}(1)}_{(1)}(x_3) \;, \cr
    p_{(1,-1)}(x_1,x_2,x_3) &= \chi^{\mr{U}(3)}_{(1,0,-1)}(x_1,x_2,x_3) + \chi^{\mr{U}(3)}_{(0,0,0)}(x_1,x_2,x_3) \;.
\end{align}
Putting everything together we find
\begin{multline}
    \chi^{\mr{U}(2)}_{(1,-1)}(x_1,x_2) = \chi^{\mr{U}(3)}_{(1,0,-1)}(x_1,x_2,x_3) + \chi^{\mr{U}(3)}_{(0,0,0)}(x_1,x_2,x_3) \\ - \chi^{\mr{U}(3)}_{(1,0,0)}(x_1,x_2,x_3)\; \chi^{\mr{U}(1)}_{(-1)}(x_3)  - \chi^{\mr{U}(3)}_{(0,0,-1)}(x_1,x_2,x_3)\; \chi^{\mr{U}(1)}_{(1)}(x_3) \;. 
\end{multline}
This leads us to conclude that, under the duality,
\begin{align}
    L^{\mr{U}(2)}_{(1,-1)} & \mapsto \l(\chi^{\mr{U}(3)}_{(1,0,-1)} + 1\r) - \chi^{\mr{U}(3)}_{(1,0,0)}\; L^{\mr{U}(1)}_{(-1)} - \chi^{\mr{U}(3)}_{(0,0,-1)}\; L^{\mr{U}(1)}_{(1)} \;. 
\end{align}

Below we give the explicit dictionary for the first few representations labelled by positive highest weights:
\begin{align}\label{EQ:duality_examples}
    L^{\mr{U}(N)}_{(0,\ldots,0)} & \mapsto L^{\mr{U}(N_{\mr{f}}-N)}_{(0,\ldots,0)} \;, \cr
    L^{\mr{U}(N)}_{(1,0,\ldots,0)} & \mapsto \chi^{\mr{U}(N_{\mr{f}})}_{(0,\ldots,0,-1)} - L^{\mr{U}(N_{\mr{f}}-N)}_{(0,\ldots,0,-1)} \;, \cr
    L^{\mr{U}(N)}_{(2,0,\ldots,0)} & \mapsto \chi^{\mr{U}(N_{\mr{f}})}_{(0,\ldots,0,-2)} - \chi^{\mr{U}(N_{\mr{f}})}_{(0,\ldots,0,-1)}L^{\mr{U}(N_{\mr{f}}-N)}_{(0,\ldots,0,-1)} + L^{\mr{U}(N_{\mr{f}}-N)}_{(0,\ldots,0,-1,-1)} \;, \cr
    L^{\mr{U}(N)}_{(1,1,0,\ldots,0)} & \mapsto \chi^{\mr{U}(N_{\mr{f}})}_{(0,\ldots,0,-1,-1)} - \chi^{\mr{U}(N_{\mr{f}})}_{(0,\ldots,0,-1)}L^{\mr{U}(N_{\mr{f}}-N)}_{(0,\ldots,0,-1)} + L^{\mr{U}(N_{\mr{f}}-N)}_{(0,\ldots,0,-2)} \;, \cr
    L^{\mr{U}(N)}_{(3,0,\ldots,0)} & \mapsto \chi^{\mr{U}(N_{\mr{f}})}_{(0,\ldots,0,-3)} - \chi^{\mr{U}(N_{\mr{f}})}_{(0,\ldots,0,-2)} L^{\mr{U}(N_{\mr{f}}-N)}_{(0,\ldots,0,-1)} + \chi^{\mr{U}(N_{\mr{f}})}_{(0,\ldots,0,-1)}L^{\mr{U}(N_{\mr{f}}-N)}_{(0,\ldots,0,-1,-1)} - L^{\mr{U}(N_{\mr{f}}-N)}_{(0,\ldots,0,-1,-1,-1)} \;, \cr
    L^{\mr{U}(N)}_{(2,1,0,\ldots,0)} & \mapsto \chi^{\mr{U}(N_{\mr{f}})}_{(0,\ldots,0,-1,-2)} -\l(\chi^{\mr{U}(N_{\mr{f}})}_{(0,\ldots,0,-2)} + \chi^{\mr{U}(N_{\mr{f}})}_{(0,\ldots,0,-1,-1)}\r)L^{\mr{U}(N_{\mr{f}}-N)}_{(0,\ldots,0,-1)} \cr
        &\qquad + \chi^{\mr{U}(N_{\mr{f}})}_{(0,\ldots,0,-1)} L^{\mr{U}(N_{\mr{f}}-N)}_{(0,\ldots,0,-2)} + \chi^{\mr{U}(N_{\mr{f}})}_{(0,\ldots,0,-1)} L^{\mr{U}(N_{\mr{f}}-N)}_{(0,\ldots,0,-1,-1)} - L^{\mr{U}(N_{\mr{f}}-N)}_{(0,\ldots,0,-1,-2)} \;, \cr
    L^{\mr{U}(N)}_{(1,1,1,0,\ldots,0)} & \mapsto \chi^{\mr{U}(N_{\mr{f}})}_{(0,\ldots,0,-1,-1,-1)} - \chi^{\mr{U}(N_{\mr{f}})}_{(0,\ldots,0,-1,-1)} L^{\mr{U}(N_{\mr{f}}-N)}_{(0,\ldots,0,-1)} + \chi^{\mr{U}(N_{\mr{f}})}_{(0,\ldots,0,-1)}L^{\mr{U}(N_{\mr{f}}-N)}_{(0,\ldots,0,-2)} - L^{\mr{U}(N_{\mr{f}}-N)}_{(0,\ldots,0,-3)} \;.
\end{align}
\end{widetext}
The dictionary for conjugate representations can be obtained by inverting all $x$'s.

It is immediate to check that the above maps are indeed involutions.

One might be puzzled by the fact that the above are written in terms of $\mr{U}(N_{\mr{f}})$ characters while $\v{G}_{\mr{F}}$ is actually $\mr{SU}(N_{\mr{f}})$. However, since, as mentioned above, the sum of all $\tau$'s is vanishing, every $\chi^{\mr{U}(N_{\mr{f}})}$ is secretly a $\v{G}_{\mr{F}}$ character as well.

The maps in \eqref{EQ:duality_examples} suggest that a more direct interpretation on the duality could be obtained by considering Wilson loops associated with supersymmetric connections including matter fields. Operators of this kind has been considered in relation to theories with boundaries \cite{Herbst:2008jq, Hori:2013ika}. Also, correspondence of such boundary supersymmetric connections under Seiberg-like dualities has been studied in \cite{Rennemo:2016oiu, Hori_to_appear}.

\acknowledgments
It is a pleasure to thank Lorenzo Bianchi, Giulio Bonelli, Matthew Buican, Kimyeong Lee, Michelangelo Preti, Diego Rodriguez-Gomez, Dario Rosa, Antonio Sciarappa, Alessandro Tanzini and Itamar Yaakov for useful discussions. We thank in particular Luca Griguolo for reading the manuscript and providing suggestions, and Kentaro Hori for many useful comments. The initial stage of this work was carried out at the \textit{Galileo Galilei Institute} in Florence during the workshop \textit{Supersymmetric Quantum Field Theories in the Non-perturbative Regime}. RP is grateful for the hospitality of the \textit{Kavli Institute for the Physics and Mathematics of the Universe} in Tokyo.

\appendix
\section{Geometry}
\subsection{Flat-space conventions}
In two-dimensional Euclidean space we introduce Dirac spinors
\begin{align}
\psi^a = \begin{pmatrix} \psi^+ \\ \psi^- \end{pmatrix} \;.
\end{align}
In this representation, the Clifford algebra is generated by gamma matrices
\begin{align}
(\gamma^\ms{1})\I{^a_b} = \begin{pmatrix} 0 & 1 \\ 1 & 0 \end{pmatrix} \;, \qquad (\gamma^\ms{2})\I{^a_b} = \begin{pmatrix} 0 & -\mr{i} \\ \mr{i} & 0 \end{pmatrix} \;.
\end{align}
They obey the identity
\begin{align}
(\gamma^{\ms{a}}\gamma^{\ms{b}})\I{^a_b} = \delta^{\ms{ab}}\delta\I{^a_b} + \mr{i}\varepsilon^{\ms{ab}}(\gamma^3)\I{^a_b} \;,
\end{align}
where $\varepsilon^{\ms{12}}=1$ and $\gamma^3$ is the chirality matrix given by
\begin{align}
(\gamma^3)\I{^a_b} = \begin{pmatrix} 1 & 0 \\ 0 & -1 \end{pmatrix} \;.
\end{align}

The charge conjugation matrix
\begin{align}
\mc{C}\I{_{ab}} = (\gamma^{\ms{2}})\I{^a_b}
\end{align}
defines the invariant Majorana product
\begin{align}\label{EQ:Majorana_product}
\psi\chi \equiv \psi^\mr{t}\mc{C}\chi = \psi_a\chi^a = \psi^a\mc{C}_{ab}\chi^b \;.
\end{align}

Since
\begin{align}
\mc{C}(\gamma^i)^{\mr{t}}\mc{C} = -\gamma^i \;,
\end{align}
it follows that, for anticommuting spinors,
\begin{align}
\psi\,\gamma^{i_1}\ldots\gamma^{i_n}\chi = (-1)^{n}\chi\,\gamma^{i_n}\ldots\gamma^{i_1}\psi \;,
\end{align}
where the $i$'s run from $\ms{1}$ to $3$.

We also define the chiral projectors
\begin{equation}
\mr{P}_{\pm} = \tf{1}{2}(\mathbf{I} \pm \gamma^3) \;.
\end{equation}

\subsection{Curved-space conventions}
Frame indices ($\ms{a}=1,2$) are denoted with a sans serif font and appear always as raised indices. The metric tensor $g$ is written in terms of a zweibein $e^{\ms{a}}$ as
\begin{align}
g = e^{\ms{a}} \otimes e^{\ms{a}} \;,
\end{align}
while the spin connection $\omega$ satisfies
\begin{align}
\d e^{\ms{a}} + \omega^{\ms{ab}} \wedge e^{\ms{b}} = 0 \;.
\end{align}
The action of the covariant derivative on spinors is defined as
\begin{align}
\nabla &= \d + \tf{1}{8}\omega^{\ms{ab}}[\gamma^{\ms{a}},\gamma^{\ms{b}}] \cr
&= \d + \tf{\mr{i}}{2}\omega^{\ms{12}}\gamma^3 \;.
\end{align}

\section{Supersymmetry}\label{APP:supersymmetry}
The Killing spinors $\epsilon$ and $\tilde{\epsilon}$ generating rigid supersymmetry are defined as Grassmann-even and satisfy the Killing spinor equations \eqref{EQ:Killing_spinor_equations}.

Since we are working in Euclidean signature, the supersymmetry algebra and all the fields are complexified. 
In quantising a theory, one specifies reality conditions for all bosonic fields. Spinor fields are defined as complex Dirac spinors and their product is taken as in \eqref{EQ:Majorana_product}, where the complex-conjugate components do not appear.

The supersymmetric variations on a generic background for components of the vector multiplet introduced in Section \ref{SEC:supersymmetry} read
\begin{align}\label{EQ:vector_multiplet_variations}
    \delta A^{\ms{a}} &= \tf{\mr{i}}{2}\epsilon\gamma^{\ms{a}}\tilde{\lambda}-\tf{\mr{i}}{2}\tilde{\epsilon}\gamma^{\ms{a}}\lambda \;, \cr
    \delta\sigma &= \epsilon\mr{P}_{+}\tilde{\lambda} + \tilde{\epsilon}\mr{P}_{-}\lambda \;, \cr
    \delta\tilde{\sigma} &= \epsilon\mr{P}_{-}\tilde{\lambda} + \tilde{\epsilon}\mr{P}_{+}\lambda \;, \cr
    \delta\lambda &= -\epsilon D + \mr{i}\gamma^3\epsilon\l(*F + \tf{1}{2}[\sigma,\tilde{\sigma}] \r) + \mr{i}\mr{P}_-\gamma^{\ms{a}}\mr{D}^{\ms{a}}(\epsilon\sigma) \cr
    &\quad + \mr{i}\mr{P}_+\gamma^{\ms{a}}\mr{D}^{\ms{a}}(\epsilon\tilde{\sigma}) \;, \cr
    \delta\tilde{\lambda} &= -\tilde{\epsilon}D - \mr{i}\gamma^3\tilde{\epsilon}\l(*F - \tf{1}{2}[\sigma,\tilde{\sigma}] \r) + \mr{i}\mr{P}_+\gamma^{\ms{a}}\mr{D}^{\ms{a}}(\tilde{\epsilon}\sigma) \cr
    &\quad + \mr{i}\mr{P}_-\gamma^{\ms{a}}\mr{D}^{\ms{a}}(\tilde{\epsilon}\tilde{\sigma}) \;, \cr
    \delta D &= -\tf{\mr{i}}{2}\big(\mr{D}^{\ms{a}}(\epsilon\gamma^{\ms{a}}\tilde{\lambda})+\mr{D}^{\ms{a}}(\tilde{\epsilon}\gamma^{\ms{a}}\lambda) + [\epsilon\mr{P}_-\tilde{\lambda} - \tilde{\epsilon}\mr{P}_+\lambda,\sigma] \cr
    &\quad + [\epsilon\mr{P}_+\tilde{\lambda} - \tilde{\epsilon}\mr{P}_-\lambda,\tilde{\sigma}]\big) \;.
\end{align}
Here, the covariant derivative acting on a field of R-charge $q$ is defined as
\begin{align}
    \mr{D} = \nabla - \mr{i}[A,\cdot\,] - \mr{i}qB \;.
\end{align}
Notice how the coupling with background graviphotons is absent since the vector multiplet has vanishing central charges.

\section{Characters}\label{APP:characters}
Irreducible representations of $\mr{U(N)}$ are labelled by a set of $N$ integers
\begin{align}
    \lambda_1 \geq \lambda_2 \geq \ldots \geq \lambda_N 
\end{align}
which form the highest weight $\v{\lambda}$ of the representation.

The character of such a representation is given by
\begin{multline}
    \chi_{\v{\lambda}}^{\mr{U}(N)}(x_1, \ldots, x_N) \\ = \f{a_{(\lambda_1+N-1, \lambda_2+N-2, \ldots, \lambda_N)}(x_1, \ldots, x_N)}{a_{(N-1, N-2, \ldots, 0)}(x_1, \ldots, x_N)} \;,
\end{multline}
where
\begin{align}
    a_{(\varrho_1, \ldots, \varrho_N)}(x_1, \ldots, x_N) = \det[x_i^{\varrho_j}]_{i,j=1}^N \;.
\end{align}
When all $\lambda$'s are nonnegative the character reduces to a \emph{Schur polynomial} $s_{\v{\lambda}}(x_1, \ldots, x_N)$.

Given an integer $k$ we define the \emph{power sum}
\begin{align}
    p_k(x_1, \ldots, x_N) = x_1^k + \ldots + x_N^k \;.
\end{align}
Analogously, for any set of ordered integers $\v{\lambda}$, we define
\begin{align}
    p_{\v{\lambda}}(x_1, \ldots, x_N) = \prod_i p_{\lambda_i}(x_1, \ldots, x_N) \;.
\end{align}
Power sums enjoy two simple properties,
\begin{align}
    p_{\v{\mu}} \, p_{\v{\nu}} & = p_{\v{\mu} \, \cup \, \v{\nu}} \;, \label{EQ:power_sum_union} \\
    p_{\v{\lambda}}(x_1^k, \ldots, x_N^k) & = p_{k\v{\lambda}}(x_1, \ldots, x_N) \;. \label{EQ:power_sum_scaling}
\end{align}

In the following we will use the notation
\begin{align}
    |\v{\lambda}| = \sum_i \lambda_i \;,
\end{align}
and define $\v{\lambda}_{\geq 0}$ and $\v{\lambda}_{< 0}$ to be, respectively, the set of nonnegative and negative entries of $\v{\lambda}$.

\subsection{From characters to power sums}\label{APP:from_characters_to_power_sum}
When $\v{\lambda}_{< 0} \neq \emptyset$, the associated $\mr{U}(N)$ character $\chi^{\mr{U}(N)}_{\v{\lambda}}$ is not a Schur polynomial, but rather some homogeneous rational function in the $x$'s. This can be split as 
\begin{align}\label{EQ:character_split}
    \chi_{\v{\lambda}}^{\mr{U}(N)}(x_1, \ldots, x_N) = (x_1 \ldots x_N)^{\lambda_N} \, s_{\tilde{\v{\lambda}}}(x_1, \ldots, x_N) \;,
\end{align}
in terms of a pure power and the Schur polynomial associated with the partition $\tilde{\v{\lambda}} = (\lambda_1 - \lambda_N, \ldots, \lambda_{N-1}-\lambda_N)$.

A given Schur polynomial can be decomposed in terms of positive power sums by means of the \emph{Frobenius formula}
\begin{align}\label{EQ:Frobenius_formula}
    s_{\v{\lambda}}(x_1, \ldots, x_N) = \sum_{|\v{\eta}| = |\v{\lambda}|} z_{\v{\eta}}^{-1}\, \hat{\chi}_{\v{\lambda}}^{\v{\eta}}\; p_{\v{\eta}}(x_1, \ldots, x_N) \;, 
\end{align}
where $\hat{\chi}_{\v{\lambda}}^{\v{\eta}}$ is the coefficient of the monomial $x_1^{\lambda_1+N-1}x_1^{\lambda_2+N-2} \ldots x_N^{\lambda_N}$ in the expansion of
\begin{align}
   p_{\v{\eta}}(x_1, \ldots, x_N)\prod_{1 \leq i < j \leq N}(x_i - x_j) \;,
\end{align}
and $z_{\v{\eta}} = \prod_n n^{a_n} a_n!$, with $a_n$ the number of times that $n$ appears $\v{\eta}$.

One can similarly decompose a generic $\mr{U}(N)$ character as a linear combination of power sums, simply by using the split as in \eqref{EQ:character_split} and by noting that
\begin{align}
    (x_1 \ldots x_N)^{\lambda_N} = s_{(1, \ldots, 1)}(x_1^{\lambda_N}, \ldots, x_N^{\lambda_N}) \;.
\end{align}
One can in fact apply \eqref{EQ:Frobenius_formula} twice, together with \eqref{EQ:power_sum_scaling} and \eqref{EQ:power_sum_union}.

Power sums of a finite number of variables are not all linearly independent which means that the above decomposition is in general not unique. One can impose additional constraints in the kind of power sums that can appear. In particular it turns out to be more convenient to require that $\dim(\v{\eta}) \leq N$, $|\v{\lambda}_{< 0}| \geq N\lambda_N$. 

\subsection{Manipulating power sums}\label{APP:manipulating_power_sum}
Given two disjoint sets $\{x_1,\ldots x_N\}$, $\{y_1,\ldots, y_M\}$, one has the trivial identity
\begin{multline}
    p_k(x_1, \ldots, x_N) = p_k(x_1, \ldots, x_N, y_1, \ldots, y_M) \\ - p_k(y_1, \ldots, y_M) \;.
\end{multline}
The above can be extended to products of power sums with
\begin{multline}
    p_{\v{\lambda}}(x_1, \ldots, x_N) = \sum_{\v{\lambda}_2 = \v{\lambda} \setminus \v{\lambda}_1} (-1)^{|\v{\lambda}_2|} \; p_{\v{\lambda}_2}(y_1, \ldots, y_M) \\ \times p_{\v{\lambda}_1}(x_1, \ldots, x_N, y_1, \ldots, y_M) \;.
\end{multline}

\subsection{From power sums to characters}\label{APP:from_power_sum_to_characters}
When $\v{\lambda}_{< 0} = \emptyset$, the associated power sum $p_{\v{\lambda}}$ can be written as a linear combination of Schur polynomials through the inverse Frobenius formula
\begin{align}\label{EQ:inverse_Frobenius_formula}
    p_{\v{\lambda}}(x_1, \ldots, x_N) = \sum_{|\v{\eta}| = |\v{\lambda}|} \hat{\chi}_{\v{\eta}}^{\v{\lambda}}\; s_{\v{\eta}}(x_1, \ldots, x_N) \;.
\end{align}
For every $\lambda_i \in \v{\lambda}_<$ we can write, as in \eqref{EQ:power_sum_union},
\begin{align}\label{EQ:power_sum_split}
    p_{\v{\lambda}} = p_{\v{\lambda}_{\geq 0}} \prod_i p_{\lambda_i} \;,
\end{align}
with
\begin{align}
    p_{\lambda_i}(x_1, \ldots, x_N) = \f{s_{(1, \ldots, 1, 0)}(x_1^{-\lambda_i}, \ldots, x_N^{-\lambda_i})}{(x_1 \ldots x_N)^{-\lambda_i}} \;.
\end{align}
Now, the numerator of the above can be expanded in terms of power sums with arguments $x_1, \ldots, x_N$ by using \eqref{EQ:Frobenius_formula} with \eqref{EQ:power_sum_scaling}. Therefore we have 
\begin{align}
    p_{\v{\lambda}}(x_1, \ldots, x_N) = \f{\sum_{\v{\mu}}\alpha_{\v{\mu}}\, p_{\v{\mu}}(x_1, \ldots, x_N)}{(x_1 \ldots x_N)^{-|\v{\lambda}_{< 0}|}} \;,
\end{align} 
for some coefficients $\alpha_{\v{\mu}}$ and $\v{\mu}_{< 0} = \emptyset$. The numerator will be some linear combination of Schur polynomial, as implied by \eqref{EQ:inverse_Frobenius_formula}. Therefore the above is some linear combination of $\mr{U}(N)$ characters as in \eqref{EQ:character_split}.

\newpage

\bibliographystyle{apsrev4-1}
\bibliography{bibliography}

\end{document}